%% file: main.tex
\def\notes{1}
\documentclass[11pt]{article}
\include{header}
\ifnum\notes=1
    
    \newcommand{\yynote}[1]{{\color{blue}\footnote{{\color{blue} {\bf YY:} #1}}}}
    
    \newcommand{\dpnote}[1]{{\color{orange}\footnote{{\color{orange} {\bf DP:} #1}}}}
\else

    \newcommand{\yynote}[1]{}
    \newcommand{\dpnote}[1]{}
\fi

\usetikzlibrary{arrows.meta,calc,positioning,decorations.pathreplacing}
\definecolor{strc}{RGB}{31,100,180}   % structured parts (block parities)
\definecolor{rndc}{RGB}{200,90,0}     % random parts (labels)
\definecolor{tabc}{RGB}{0,128,90}     % table values
\definecolor{sigc}{RGB}{190,30,45}    % the sign

\title{Submodularity Is Barely Testable}

\author{
    Diptaksho Palit\thanks{This project was carried out during his participation in the 2026 NII International Internship Program.} \\
    Boston University \\
    \texttt{dpalit@bu.edu}
    \and
    Yuichi Yoshida\thanks{Supported by JSPS KAKENHI Grant Numbers 25K24465 and 26K21777}\\
    National Institute of Informatics\\
    \texttt{yyoshida@nii.ac.jp}
}

\date{\today}
\begin{document}

\maketitle

\begin{abstract}
    In 2010, Seshadhri and Vondr\'ak asked: {\em is submodularity testable?}
    For functions $f : \{ 0, 1 \}^n \to \mathbb{R}$, they gave a tester that makes $2^{O(\sqrt{n} \log n)}$ queries.
    We show that every tester, even an adaptive one with two-sided error, must make $2^{\Omega(\sqrt{n})}$ queries, thereby concluding: {\em submodularity is barely testable}.
\end{abstract}

% \newpage
% \setcounter{tocdepth}{2}
% \tableofcontents

\newpage

\section{Introduction}
\label{sec:intro}

A set function $f\colon 2^{[n]}\to\mathbb{R}$ is \emph{submodular} if
\[
f(S)+f(T)\;\ge\; f(S\cup T)+f(S\cap T)\qquad\text{for all } S,T\subseteq[n].
\]
Equivalently, the marginal value $f(S\cup\{i\})-f(S)$ of an element $i$ can only decrease as $S$ grows.
% This principle of diminishing returns makes submodular functions ubiquitous.
Submodular functions are a discrete analog of convex functions in combinatorial optimization~\cite{Edmonds01,NemhauserWF78,Lovasz82,Schrijver00,IwataFF01}, they model the valuations of buyers in algorithmic game theory~\cite{LehmannLN02,DobzinskiS06,Vondrak08}, and they measure the information collected by a set of sensors in machine learning~\cite{GuestrinKS05,KrauseGGK06,KrauseSG08}.
 
Algorithms for submodular functions usually access the function through a value oracle.
Such an algorithm generally runs in time polynomial in $n$, so it sees only a tiny fraction of the $2^n$ values, even though its guarantees rely on a global property of the function. % \yynote{Could we restrict this statement to the efficient algorithms motivating the discussion? Value-oracle access alone does not imply polynomial running time.}
% This makes it a natural target for property testing \cite{GGR98}.
Seshadhri and Vondr\'ak~\cite{SeshadhriV14} therefore asked whether one can efficiently test if a function is submodular.
% This is a question about property testing~\cite{RS96,GGR98}.
In the context of property testing \cite{GoldreichGR98}, a function is \emph{$\varepsilon$-far} from submodularity if at least an $\varepsilon$ fraction of its values must be changed to make it submodular, that is, if its normalized Hamming distance to every submodular function is at least $\varepsilon$.
An \emph{$\varepsilon$-tester} queries the function at points of its choice.
It must accept every submodular function and reject every function that is $\varepsilon$-far from submodular, each with probability at least $2/3$.
We give the formal definitions in \cref{sec:prelims}.
 
Seshadhri and Vondr\'ak showed that submodularity can be tested with far fewer than $2^n$ queries.
Their tester samples $(1/\varepsilon)^{O(\sqrt{n}\log n)}$ random squares $\{ S, \, S \cup \{ i \} , \, S \cup \{j\} , \, S \cup \{i, j \} \}$ and checks whether the \emph{slack} of the square,
\[
    f(S \cup \{ i \}) + f(S \cup \{ j \}) - f(S) - f(S \cup \{ i, j \}),
\]
is negative, rejecting if so.
This is an equivalent definition of submodularity: every square must have nonnegative slack (\cref{lem:squares}).
The tester is nonadaptive and has one-sided error. % \yynote{Could we define slack somewhere and state that nonnegative slack on every square is equivalent to submodularity? The overview uses both before the preliminaries.}
This query complexity is subexponential, but far from polynomial, and they asked whether there is a tester with $\mathrm{poly}(n/\varepsilon)$ queries, or with $\mathrm{poly}(n)$ queries for constant $\varepsilon$; perhaps an adaptive one or one that searches for obstructions other than violated squares~\cite[Section~6]{SeshadhriV14}.
% The known lower bounds are only polynomial.
% Seshadhri and Vondr\'ak also reduced monotonicity testing to submodularity testing.
The best known lower bound for this problem was only polynomial, due to  Hatami and Vondr\'ak~\cite{HatamiV16}, who showed that every (even adaptive, two-sided error) tester must make $\Omega(n^2/\log n)$ queries.
For constant $\varepsilon$, the query complexity of testing submodularity was thus known only to lie between $\Omega(n^2/\log n)$ and $2^{O(\sqrt{n}\log n)}$, and resolving this gap has been a major open question in property testing \cite{Sublinear37}.
 
We show that the answer to the question of Seshadhri and Vondr\'ak is negative, and that their tester is nearly optimal.
\begin{theorem}
    \label{thm:main}
    For $\varepsilon \in \left( 0, \frac{1}{10^8} \right)$, every $\varepsilon$-tester for submodularity must make at least $2^{\sqrt{n}/5}$ queries on functions from $\{ 0, 1 \}^n$ to $\mathbb{R}$.
\end{theorem}
 
This bound holds even for adaptive testers with two-sided error.
For constant $\varepsilon$, it matches the upper bound of Seshadhri and Vondr\'ak up to a logarithmic factor in the exponent.
% \yynote{Could we explain what resolving this question teaches us about submodularity or property testing, beyond closing the quantitative gap (if any)? \dipi{We can, I just don't know what to say lol. You can add whatever you think is necessary.}}
% Submodularity is therefore barely testable: it can be tested with a number of queries that is sublinear in the size of the domain, but not with a number of queries that is polynomial in $n$. The theorem also separates submodularity from monotonicity. By the reduction above, testing submodularity is at least as hard as testing monotonicity, and monotonicity can be tested with $O(n^2/\varepsilon)$ queries~\cite{DGLRRS99}, so testing submodularity is in fact superpolynomially harder.
 
% The functions in our construction take values that are integer multiples of $2^{-n}$ with absolute value at most $n^3$. An increasing affine map preserves submodularity and changes no distances, so the lower bound also holds for functions with range $[0,1]$ whose values have $O(n)$ bits of precision.
% The bound |F| <= n^3 holds for m >= 3: the minimum is at slice 11 with
% X = A, Y = B, where |F| <= 4m(m^2+m)^2 + 4m^6 + m <= 8m^6 <= n^3, and
% every value is below m + 1. Section 3 does not state this yet; consider
% adding a one-line remark after the definition of F^(sigma).
 
\subsection{Related work}
 
\paragraph{Testing submodularity in Hamming distance.}
Parnas, Ron and Rubinfeld~\cite{ParnasRR02} first raised the problem of testing submodularity, for functions on grids.
They focused on two-dimensional grids, where submodular functions correspond to Monge matrices.
Seshadhri and Vondr\'ak~\cite{SeshadhriV14} initiated the study of testing submodularity on the hypercube.
Besides their tester, they showed that the obstructions to submodularity can be complex.
For every $\varepsilon\ge 2^{-n/10}$, there is a function that is $\varepsilon$-far from submodular whose density of violated squares is less than $\varepsilon^{4.8}$.
There are also partial functions on exponentially many points that have no submodular extension, although every proper restriction of them has one.
Bhaskar and Kumar~\cite{BhaskarK20} studied such extension problems further.
On the lower-bound side, a reduction from monotonicity testing~\cite{SeshadhriV14} combined with the lower bounds of Fischer, Lehman, Newman, Raskhodnikova, Rubinfeld, and Samorodnitsky~\cite{FischerLNRRS02} gives lower bounds of $\Omega(\sqrt{n})$ queries for nonadaptive one-sided testers and $\Omega(\log n)$ queries for adaptive one-sided testers.
The lower bound for general testers was later improved to $\Omega(n)$~\cite{BlaisBM12,RaskhodnikovaY13} and then to $\Omega(n^2/\log n)$~\cite{HatamiV16}.
 
\paragraph{Bounded ranges and special classes.}
Raskhodnikova and Yaroslavtsev~\cite{RaskhodnikovaY13} showed that every submodular function $f\colon\{0,1\}^n\to\{0,1,\dots,k\}$ can be represented as a pseudo-Boolean $2k$-DNF.
Their learning algorithm for such representations gives a tester for submodularity of these functions that makes $\mathrm{poly}(n)$ queries when $k=O\bigl((\log n/\log\log n)^{1/2}\bigr)$ and $\varepsilon$ is constant.
Chakrabarty and Huang~\cite{ChakrabartyH15} gave an exact learning algorithm for coverage functions, a subclass of submodular functions, which also yields a tester for this subclass~\cite{RaskhodnikovaY13}.
% These results do not apply to our construction, whose functions take superpolynomially many distinct values.\yynote{Coverage functions can also take exponentially many distinct values. Could we avoid using the number of values as a reason to exclude the coverage-testing result?}
More recently, Haris and Palit~\cite{HarisP26} studied testing $k$-submodularity, a generalization of submodularity to partial partitions of the ground set that has been studied in optimization~\cite{WardZ14,OhsakaY15}.
They gave a constant-query tester in $\ell_p$ distance, subexponential-query testers in Hamming distance for the two local conditions that together characterize $k$-submodularity, and a tester for monotone $k$-submodular functions with bounded integer range. % \yynote{Could we say ``bounded integer range''? Their Theorem 1.5 assumes range $\{0,1,\ldots,r\}$, which is different from merely taking real values in $[0,1]$.}
For $k=1$, $k$-submodularity is ordinary submodularity, so our lower bound applies to this case.
 
\paragraph{Testing in $\ell_p$ distance.}
Berman, Raskhodnikova and Yaroslavtsev~\cite{BermanRY14} introduced $\ell_p$-testing, which measures the distance between functions $f,g\colon\{0,1\}^n\to[0,1]$ by $\|f-g\|_p$ instead of the fraction of points where they differ.
Feldman and Vondr\'ak~\cite{FeldmanV16} showed that every submodular function $f\colon\{0,1\}^n\to[0,1]$ is $\varepsilon$-close in $\ell_2$ distance to a function of $O(\varepsilon^{-2}\log(1/\varepsilon))$ variables, and used this to test submodularity in $\ell_1$ distance with $\mathrm{poly}(1/\varepsilon)\log n+2^{\tilde O(1/\varepsilon)}$ queries.
Blais and Bommireddi~\cite{BlaisB17} extended the testing-by-implicit-learning framework~\cite{DiakonikolasLMORSW07} to obtain, for every $p\ge 1$, $\ell_p$-testers whose number of queries is independent of $n$, for submodularity and for several other classes of valuation functions.
Our lower bound is consistent with these results.
In our construction, a YES function and a NO function with the same random choices differ by exactly $2^{1-n}$ on a quarter of the domain and agree elsewhere. So every NO function is exponentially close to a submodular function in every $\ell_p$ distance, even though it is (with high probability) far from submodularity in Hamming distance. % \yynote{The first assertion holds for every NO function, but the Hamming-distance assertion holds only with probability at least $8/10$. Could we state these guarantees separately? There really are submodular functions in the NO support: the allowed realization $\phi_A(X)=\pi_A(X)$ and $\phi_B(Y)=\pi_B(Y)\oplus e_1$ has no centers, so Lemma~\ref{lem:square-slack} makes $F^{(-1)}$ submodular.}
 
\paragraph{Learning submodular functions.}
Testing and learning are closely related~\cite{GoldreichGR98}, and several of the testers above come from learning algorithms~\cite{ChakrabartyH15,RaskhodnikovaY13,FeldmanV16,BlaisB17}.
Learning submodular functions and related classes of valuation functions has been studied extensively~\cite{GoemansHIM09,BalcanH12,GuptaHRU13,CheraghchiKKL12,BalcanCIW12,BadanidiyuruDFKNR12,FeldmanKV13,RaskhodnikovaY13,FeldmanV16}.
For example, Goemans, Harvey, Iwata and Mirrokni~\cite{GoemansHIM09} showed how to approximate a nonnegative monotone submodular function everywhere within a factor of $\tilde O(\sqrt{n})$ using polynomially many value queries, % \yynote{Could we add ``nonnegative'' to ``monotone submodular function''? This is a hypothesis of the multiplicative approximation result; our standing real-valued setting does not imply it.}
and Balcan and Harvey~\cite{BalcanH12} introduced the PMAC model for learning submodular functions from random examples.
A learning algorithm may assume that its input is submodular, whereas a tester must also handle inputs that are far from submodular~\cite{SeshadhriV14}.

\subsection{Open questions}
\label{sec:open}
 
\paragraph{The logarithmic factor.}
For constant $\varepsilon$, the query complexity of testing submodularity lies between $2^{\Omega(\sqrt{n})}$ and $2^{O(\sqrt{n}\log n)}$.
Resolving this log-factor is still open.
 
\paragraph{Bounded ranges.}
Our hard functions take superpolynomially many distinct values, whereas submodularity of functions into $\{0,1,\dots,k\}$ can be tested with $\mathrm{poly}(n)$ queries for constant $\varepsilon$ when $k=O\bigl((\log n/\log\log n)^{1/2}\bigr)$~\cite{RaskhodnikovaY13}.
At which range size does testing become superpolynomially hard? For example, can functions with range size $\mathrm{poly}(n)$ be tested with $\mathrm{poly}(n)$ queries?
 
\paragraph{Lower bounds for $k$-submodularity.}
For $k\ge 2$, no nontrivial lower bound is known for testing $k$-submodularity: lower bounds for submodularity do not transfer directly, since natural embeddings can destroy pairwise monotonicity~\cite{HarisP26}.
A nontrivial lower bound for testing $k$-submodularity or even monotone $k$-submodularity would be interesting.

\subsection{Technical overview}
 
We construct two distributions over functions $\{0,1\}^n\to\mathbb{R}$. Every function in the YES distribution is submodular, a function in the NO distribution is far from submodular with high probability, and an algorithm that makes fewer than $2^{\Theta(\sqrt{n})}$ queries receives almost the same answers under both. This overview explains where each part of the construction comes from. We start from two results of Seshadhri and Vondr\'ak~\cite{SeshadhriV14} that shaped our goals, then add the components of the construction one at a time, and finally sketch the two halves of the analysis.
 
\paragraph{What a hard instance must look like.}
A NO function must have violated squares, that is, squares with negative slack, and an algorithm that finds one knows that the function is not submodular. 
The tester of Seshadhri and Vondr\'ak samples random squares, and its analysis shows how rare violated squares can be.
They repair a violated square by lowering the function, by the amount of the violation, on all points below the square, or on all points above it; this fixes the square without creating new violations~\cite[Lemma~2.2]{SeshadhriV14}.
If there are at most $\varepsilon^{\sqrt{n}\log n}2^n$ violated squares, these repairs change at most $\varepsilon 2^n$ values in total~\cite[Lemma~2.3]{SeshadhriV14}.
So for constant $\varepsilon$, a function that is $\varepsilon$-far from submodular has at least a $2^{-O(\sqrt{n}\log n)}$ fraction of violated squares.
 
The violated squares of a hard NO function must therefore form at least a $2^{-O(\sqrt{n}\log n)}$ fraction of all squares.
We aim for a $2^{-\Theta(\sqrt{n})}$ fraction, at random positions, so that an algorithm needs about $2^{\Theta(\sqrt{n})}$ queries before it sees one.
% The repair strategy above suggests that this is the right regime.
% If violated squares this sparse are spread out at random near the middle of the cube, the down-sets and up-sets that the strategy lowers already cover a constant fraction of the cube.\yynote{Could we drop the repeated density bound and keep the repair argument as motivation? A costly repair strategy alone does not establish distance from submodularity.}
 
\paragraph{One violated square that forces many changes.}
Our starting point is a function of Seshadhri and Vondr\'ak with only one violated square, which nevertheless needs many changes to become submodular~\cite[Section~3.2]{SeshadhriV14}.
It has two special coordinates, which we call \emph{action coordinates}; the remaining $k:=n-2$ coordinates are \emph{control coordinates}.
Fixing the action coordinates to $\tau\in\{00,01,10,11\}$ gives a copy of $\{0,1\}^k$, which we call the \emph{slice} $\tau$. For a fixed $z\in\{0,1\}^k$ with $|z|=k/2$, the function is
\[
    f(x,00)=|x|,\qquad f(x,01)=f(x,10)=|x\oplus z|,\qquad f(x,11)=1-|x|.
\]
The square spanned by the two action coordinates at $x$ has slack $f(x,01)+f(x,10)-f(x,00)-f(x,11)=2|x\oplus z|-1$, so it is violated exactly when $x=z$, and every other square has nonnegative slack.
 
To see why $f$ needs many changes, let $G$ be submodular.
Then $G$ differs from $f$ at one of the four points $(z,\tau)$.
In that slice, $f$ is linear in $x$, so $g:=G(\cdot,\tau)-f(\cdot,\tau)$ is submodular, and $g(z)\neq0$.
A submodular function cannot have a lone nonzero value in the middle of the cube.
If $g(z)>0$, then $g$ is nonzero on at least half of the subcube above $z$~\cite[Lemma~3.5]{SeshadhriV14}; if $g(z)<0$, then $g$ is nonzero at no fewer than $2^{k/2}$ points~\cite[Lemma~3.6]{SeshadhriV14}.
So repairs \emph{propagate}: the single violated square forces $\Omega(2^{k/2})$ changes.
 
This example has two shortcomings.
First, $2^{k/2}$ changes are an exponentially small fraction of the domain; we need many violated squares whose forced changes do not overlap much.
Second, the violated square is easy to find: the values $f(0,01)$ and $f(e_i,01)$ for $i\in[k]$ reveal $z$.
Our construction keeps the skeleton of this example (two action coordinates, slices on which the function is linear on suitable subcubes, and a distance term that vanishes exactly at the violated squares) and fixes both shortcomings. % \yynote{Could we replace ``close to linear'' by ``modular on suitable subcubes''? No approximation to a linear function is used.}
 
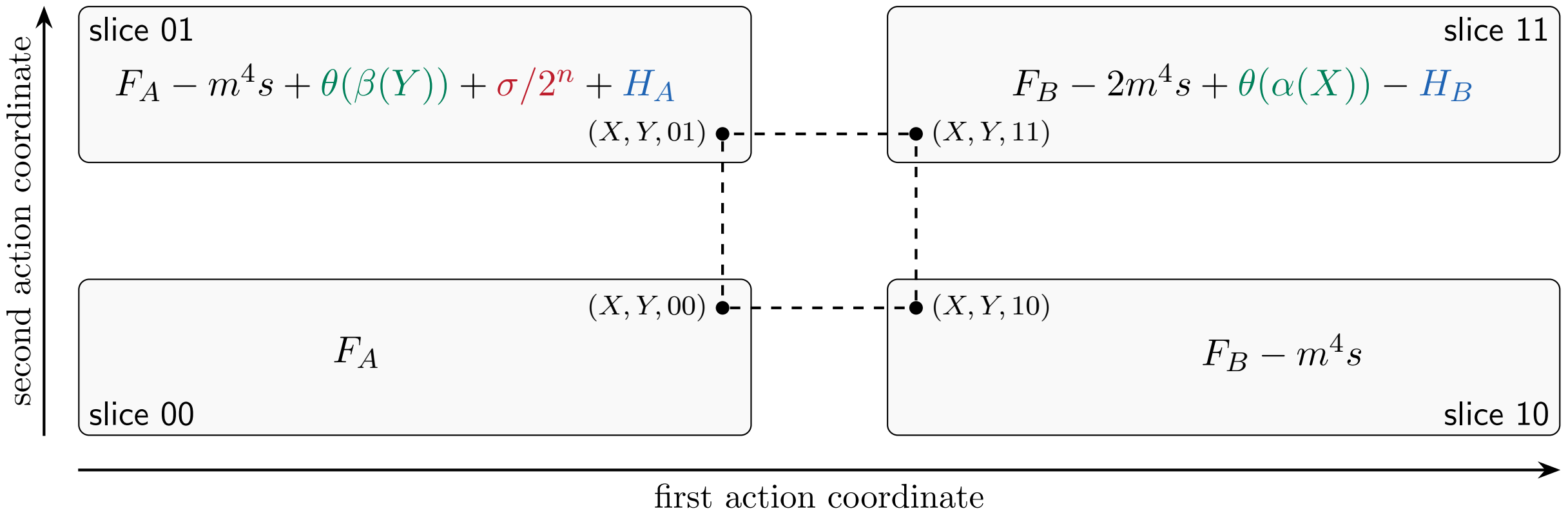
\begin{figure}[t]
\centering
\begin{tikzpicture}[
  slice/.style={draw, rounded corners=3pt, minimum width=6.9cm, minimum height=1.6cm, fill=gray!5},
  lab/.style={font=\small\sffamily, inner sep=3pt},
  dot/.style={circle, fill=black, inner sep=1.4pt},
  pl/.style={font=\scriptsize, inner sep=2pt}
]
\node[slice] (s00) at (0,0) {};
\node[slice] (s10) at (8.3,0) {};
\node[slice] (s01) at (0,2.8) {};
\node[slice] (s11) at (8.3,2.8) {};
\node[lab, anchor=south west] at (s00.south west) {slice 00};
\node[lab, anchor=south east] at (s10.south east) {slice 10};
\node[lab, anchor=north west] at (s01.north west) {slice 01};
\node[lab, anchor=north east] at (s11.north east) {slice 11};
\node at ($(s00.center)+(-0.6,0.05)$) {$F_A$};
\node at ($(s10.center)+(0.6,0.05)$) {$F_B-m^4s$};
\node at ($(s01.center)+(-0.2,0)$) {$F_A-m^4s+\textcolor{tabc}{\theta(\beta(Y))}+\textcolor{sigc}{\sigma/2^n}+\textcolor{strc}{H_A}$};
\node at ($(s11.center)+(0.2,0)$) {$F_B-2m^4s+\textcolor{tabc}{\theta(\alpha(X))}-\textcolor{strc}{H_B}$};
\node[dot] (d00) at ($(s00.north east)+(-0.3,-0.3)$) {};
\node[dot] (d10) at ($(s10.north west)+(0.3,-0.3)$) {};
\node[dot] (d01) at ($(s01.south east)+(-0.3,0.3)$) {};
\node[dot] (d11) at ($(s11.south west)+(0.3,0.3)$) {};
\node[pl, anchor=east] at (d00.west) {$(X,Y,00)$};
\node[pl, anchor=west] at (d10.east) {$(X,Y,10)$};
\node[pl, anchor=east] at (d01.west) {$(X,Y,01)$};
\node[pl, anchor=west] at (d11.east) {$(X,Y,11)$};
\draw[thick, dashed] (d00) -- (d10) -- (d11) -- (d01) -- (d00);
\draw[-{Stealth}, thick] ($(s00.south west)+(0,-0.35)$) -- node[below, font=\small] {first action coordinate} ($(s10.south east)+(0,-0.35)$);
\draw[-{Stealth}, thick] ($(s00.south west)+(-0.35,0)$) -- node[above, sloped, font=\small] {second action coordinate} ($(s01.north west)+(-0.35,0)$);
\end{tikzpicture}
\caption{The function $F^{(\sigma)}$ on its four slices, where $s:=|X|+|Y|$. The dots are the four points $(X,Y,\tau)$, which form the square spanned by the two action coordinates at $(X,Y)$. Its slack is $F^{(\sigma)}(X,Y,01)+F^{(\sigma)}(X,Y,10)-F^{(\sigma)}(X,Y,00)-F^{(\sigma)}(X,Y,11)=\textcolor{strc}{H_A+H_B}+\textcolor{tabc}{\theta(\beta(Y))-\theta(\alpha(X))}+\textcolor{sigc}{\sigma/2^n}$, since all other terms cancel.}
\label{fig:slices}
\end{figure}
 
\paragraph{Hiding the violated squares with addresses.}
To hide the violated squares, we want the distance term to vanish on a random set of points that queries cannot reveal.
This seems to clash with the distance argument, which needs linear slices: a slice that is linear in all $k$ control coordinates is determined by $k+1$ of its values, so its zeros can be learned, as in the example.
But the argument propagates a change within a single slice, so each slice needs to be linear only in the coordinates along which the change propagates. 
This is why we split the control coordinates into two halves $A$ and $B$; we write points as $(X,Y,\tau)$ with $X\subseteq A$, $Y\subseteq B$ and $\tau$ the slice.
In slices $00$ and $01$, we propagate changes by varying $X$, so these slices may depend on $Y$ arbitrarily, and in particular randomly; in slices $10$ and $11$, the roles of $X$ and $Y$ are exchanged.
Since the square spanned by the action coordinates at $(X,Y)$ involves all four slices, whether it is violated can depend randomly on both halves.
 
We give every $X$ an \emph{address} $\alpha(X)\in\{0,1\}^{2m}$ and every $Y$ an address $\beta(Y)\in\{0,1\}^{2m}$, and we replace $|x\oplus z|$ by the Hamming distance $|\alpha(X)\oplus\beta(Y)|$ between the two addresses.
The square spanned by the action coordinates at $(X,Y)$ is then violated exactly when $\alpha(X)=\beta(Y)$; we call such pairs \emph{centers}.
If the addresses were uniformly random, the centers would form a random $2^{-2m}$ fraction of all pairs, and an algorithm could not locate them.

\begin{figure}[t]
\centering
\begin{tikzpicture}[
  part/.style={draw, minimum height=0.75cm, minimum width=2.6cm, font=\small},
  cell/.style={draw, minimum size=0.28cm, inner sep=0pt},
  brace/.style={decorate, decoration={brace, amplitude=4pt}}
]
% coordinate layout
\node[font=\small, anchor=east] at (-0.2,2.55) {coordinates:};
\foreach \i in {0,1} \node[cell, fill=gray!40] at (0.2+0.3*\i,2.55) {};
\foreach \b/\x in {1/1.3, 2/2.6, m/4.55} {
  \foreach \j in {0,...,3} \node[cell, fill=white] at (\x+0.3*\j,2.55) {};
}
\node at (3.95,2.55) {$\cdots$};
\foreach \b/\x in {1/6.3, 2/7.6, m/9.55} {
  \foreach \j in {0,...,3} \node[cell, fill=white] at (\x+0.3*\j,2.55) {};
}
\node at (8.95,2.55) {$\cdots$};
\draw[brace] (0.05,2.8) -- node[above=4pt, font=\scriptsize] {action} (0.65,2.8);
\draw[brace] (1.15,2.8) -- node[above=4pt, font=\scriptsize] {$A_1$} (2.35,2.8);
\draw[brace] (2.45,2.8) -- node[above=4pt, font=\scriptsize] {$A_2$} (3.65,2.8);
\draw[brace] (4.4,2.8) -- node[above=4pt, font=\scriptsize] {$A_m$} (5.6,2.8);
\draw[brace] (6.15,2.8) -- node[above=4pt, font=\scriptsize] {$B_1$} (7.35,2.8);
\draw[brace] (7.45,2.8) -- node[above=4pt, font=\scriptsize] {$B_2$} (8.65,2.8);
\draw[brace] (9.4,2.8) -- node[above=4pt, font=\scriptsize] {$B_m$} (10.6,2.8);
\draw[brace, decoration={mirror}] (1.15,2.3) -- node[below=4pt, font=\scriptsize] {half $A$: contains $X$} (5.6,2.3);
\draw[brace, decoration={mirror}] (6.15,2.3) -- node[below=4pt, font=\scriptsize] {half $B$: contains $Y$} (10.6,2.3);
% addresses
\node[font=\small, anchor=east] at (1.6,0.75) {$\alpha(X)=$};
\node[part, fill=strc!15] (pa) at (3.0,0.75) {$\pi_A(X)$};
\node[part, fill=rndc!15, right=0pt of pa] (fa) {$\phi_A(X)$};
\node[font=\small, anchor=east] at (1.6,-1.25) {$\beta(Y)=$};
\node[part, fill=rndc!15] (fb) at (3.0,-1.25) {$\phi_B(Y)$};
\node[part, fill=strc!15, right=0pt of fb] (pb) {$\pi_B(Y)$};
\draw[{Stealth}-{Stealth}, thick] (pa.south) -- node[left, font=\scriptsize, align=right] {$H_A$\\(slice 01)} (fb.north);
\draw[{Stealth}-{Stealth}, thick] (fa.south) -- node[right, font=\scriptsize, align=left] {$H_B$\\(slice 11)} (pb.north);
% legend
\node[draw, fill=strc!15, minimum size=0.3cm] (l1) at (7.3,0.35) {};
\node[anchor=west, font=\scriptsize, align=left] at (l1.east) {structured: one parity bit\\per block of its own half};
\node[draw, fill=rndc!15, minimum size=0.3cm] (l2) at (7.3,-0.75) {};
\node[anchor=west, font=\scriptsize, align=left] at (l2.east) {random: a uniform label,\\drawn independently for every set};
\end{tikzpicture}
\caption{Top: the coordinates (shown with blocks of four). Bottom: the addresses. $(X,Y)$ is a center when both pairs joined by arrows are equal, which happens with probability $2^{-2m}$. Each comparison involves the structured part of one half and the random part of the other, so for fixed $Y$ the term $H_A$ in slice $01$ is a sum over the blocks of $X$, and for fixed $X$ the term $H_B$ in slice $11$ is a sum over the blocks of $Y$.}
\label{fig:addresses}
\end{figure}
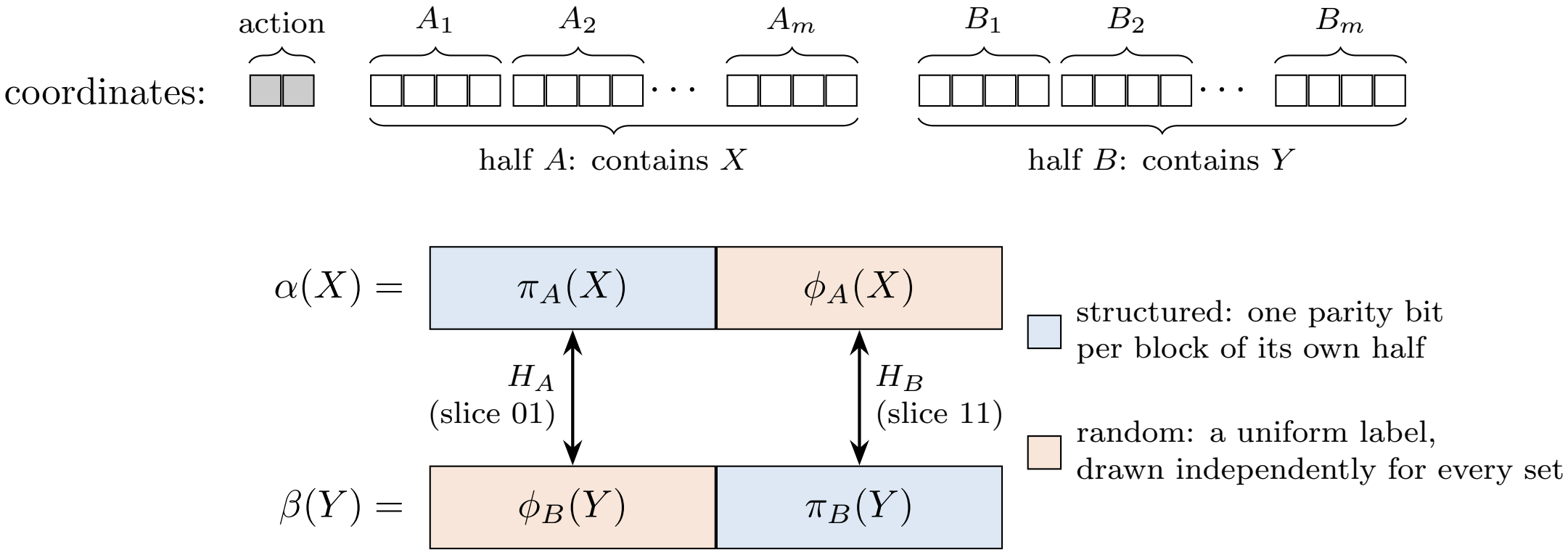

However, uniformly random addresses would make the distance term random in both halves, while every slice that contains it must be structured in one of them.
% However, the distance argument above needs each slice to be linear, and a random address is far from linear.
We therefore make each address half structured and half random.
We divide each half into $m$ blocks $A_1,\dots,A_m$ and $B_1,\dots,B_m$ of $m$ coordinates each (so that $n = 2m^2 + 2$), % \yynote{Could we state $n=2m^2+2$ here, connecting the address length to the $\sqrt n$ scale?}
and let
\[
    \alpha(X):=\bigl(\textcolor{strc}{\pi_A(X)},\,\textcolor{rndc}{\phi_A(X)}\bigr),\qquad
    \beta(Y):=\bigl(\textcolor{rndc}{\phi_B(Y)},\,\textcolor{strc}{\pi_B(Y)}\bigr).
\]
Here the structured part $\pi_A(X)\in\{0,1\}^m$ lists the parities of the blocks $X\cap A_i$, and the random part $\phi_A(X)\in\{0,1\}^m$ is a uniformly random \emph{label}, drawn independently for every set $X$; $\beta(Y)$ is defined in the same way.
Note the order: each part of $\alpha(X)$ is compared with the part of $\beta(Y)$ of the other kind.
As a result, the distance splits into two terms,
\[
    |\alpha(X)\oplus\beta(Y)| = H_A(X,Y)+H_B(X,Y),\qquad
    H_A:=|\textcolor{strc}{\pi_A(X)}\oplus\textcolor{rndc}{\phi_B(Y)}|,\quad
    H_B:=|\textcolor{rndc}{\phi_A(X)}\oplus\textcolor{strc}{\pi_B(Y)}|,
\]
each of which compares a structured part of one half with a random part of the other (Figure~\ref{fig:addresses}).
We place $H_A$ in slice $01$ and $-H_B$ in slice $11$, so that the slack gains $H_A+H_B$ (Figure~\ref{fig:slices}).
For a fixed $Y$, the term $H_A$ is a sum of $m$ terms, each depending only on the parity of one block $X\cap A_i$. 
Symmetrically, for a fixed $X$, the term $H_B$ is a sum of terms each depending only on one block $Y\cap B_i$.
 
Every component we add keeps this property: in slices $00$ and $01$, for each fixed $Y$, the function is a sum of functions of single blocks $X\cap A_i$, and in slices $10$ and $11$ the same holds with the roles of $X$ and $Y$ exchanged.
This is weaker than linearity, but it is enough for the distance argument.
On every subcube in which exactly one coordinate per block is free, a sum of functions of single blocks is modular, and modularity is all the argument needs.
 
At the same time, the centers stay hidden.
A pair $(X,Y)$ is a center when the parities of $X$ equal the label of $Y$ and the label of $X$ equals the parities of $Y$. 
Labels are independent of everything else, so an algorithm knows nothing about the label of a set before querying it.
 
\paragraph{Hiding the sign with a shared random table.}
The YES and NO functions differ only in the sign $\sigma\in\{+1,-1\}$ of the slack at centers.
We cannot simply add $\sigma/2^n$ to slice $01$: all other terms are integers, so the fractional part of a single answer would reveal $\sigma$.
Instead, we sample a table $\theta$ that assigns an independent uniform value in $\{0,\frac{1}{2^n},\dots,\frac{2^n-1}{2^n}\}$ to every address.
We add $\theta(\beta(Y))+\sigma/2^n$ to slice $01$ and $\theta(\alpha(X))$ to slice $11$.
The slack of the square spanned by the action coordinates at $(X,Y)$ becomes
\[
    |\alpha(X)\oplus\beta(Y)|+\theta(\beta(Y))-\theta(\alpha(X))+\frac{\sigma}{2^n}.
\]
At a center, the two table values coincide, and the slack is exactly $\sigma/2^n$: positive for YES and negative for NO. 
Away from centers, the distance term is at least $1$, while the table values and the sign together contribute at least $-1$, so the slack is nonnegative (\cref{clm:action}).
A query in slice $01$ reveals $\theta(\beta(Y))+\sigma/2^n$, a uniform value on a grid shifted by $\sigma/2^n$, which says nothing about $\sigma$ on its own.
The shift becomes visible only if the algorithm also sees $\theta(\beta(Y))$ unshifted, which happens only in slice $11$ at a set $X$ with $\alpha(X)=\beta(Y)$, that is, after it has found a center.
 
\paragraph{Keeping the other squares submodular.}
The random terms can create violated squares spanned by two control coordinates. 
In slice $01$, for example, $\theta(\beta(Y))+H_A(X,Y)$ depends on $Y$ through the random label $\phi_B(Y)$, so it behaves like an arbitrary function of $Y$ with values in $[0,m+1)$.
We absorb these fluctuations with the \emph{base components}
\[
    F_A(X,Y):=-4\sum_{i=1}^m\bigl(|X\cap A_i|+|Y|\bigr)^2,\qquad F_B(X,Y):=-4\sum_{i=1}^m\bigl(|Y\cap B_i|+|X|\bigr)^2,
\]
using $F_A$ in slices $00$ and $01$, and $F_B$ in slices $10$ and $11$.
Each summand is a concave function of a count.
So $F_A$ has slack $8m$ on squares spanned by two coordinates of $B$, and slack $8$ on squares spanned by two coordinates of one block of $A$, or by one coordinate from each half (\cref{clm:base}).
This is more than the random terms can take away.
Only squares spanned by coordinates of two different blocks of $A$ get slack $0$ from $F_A$, and there the random terms are sums over blocks as well, so they contribute slack $0$ too (\cref{clm:control}). % \yynote{The random-term argument is in Claim~\ref{clm:control}.}
This is the first role of the blocks.
A single concave function such as a large enough multiple of $-(|X|+|Y|)^2$ would also absorb the random terms, but it is not a sum of functions of single blocks of $X$, so the restrictions used in the distance argument would no longer be modular. % \yynote{Could we use a sufficiently large multiple, e.g.\ $-4m(|X|+|Y|)^2$? The displayed quadratic has slack only $2$, whereas the random terms can contribute $-2m$.}
 
Squares spanned by one action coordinate and one control coordinate need one more term.
The slack of such a square is the decrease, when a control element is added, of the difference between two neighboring slices. 
These differences contain $F_B-F_A$ and the random terms, which can grow by $O(m^3)$ when one element is added (\cref{clm:growth}).
We therefore subtract the \emph{linear correction} $m^4(|X|+|Y|)$ once in slices $01$ and $10$ and twice in slice $11$.  
Every difference between neighboring slices then drops by $m^4$ whenever an element is added, which outweighs any growth of the other terms (\cref{clm:mixed}).
The linear terms have slack $0$ on every square spanned by control coordinates, and they cancel on every square spanned by action coordinates, so the computations above stay intact. % , and they cancel in the square spanned by the action coordinates.\yynote{The correction is linear only within each slice. Could we say that it leaves control--control and action--action slacks unchanged, but adds $m^4$ to mixed slacks?}
 
\begin{figure}[t]
\centering
\begin{tikzpicture}[
  cell/.style={draw, minimum size=0.36cm, inner sep=0pt},
  pt/.style={draw, rounded corners=2pt, fill=white, font=\small, inner sep=3pt}
]
% left panel: blocks of A as columns
\foreach \c/\x/\lab in {1/0/A_1, 2/0.75/A_2, 3/1.5/A_3, 4/2.6/A_m} {
  \foreach \r in {0,...,5} \node[cell] (c\c r\r) at (\x,0.4*\r) {};
  \node[font=\small] at (\x,-0.5) {$\lab$};
}
\node at (2.05,1.0) {$\cdots$};
% X: shaded cells
\foreach \c/\r in {1/0,1/1,1/3, 2/1,2/2,2/4, 3/0,3/2,3/5, 4/1,4/3,4/4} \fill[gray!35] (c\c r\r.south west) rectangle (c\c r\r.north east);
\foreach \c in {1,2,3,4} \foreach \r in {0,...,5} \draw (c\c r\r.south west) rectangle (c\c r\r.north east);
% Gamma: one star per block
\foreach \c/\r in {1/3, 2/0, 3/4, 4/4} \node[font=\small] at (c\c r\r) {$\star$};
\node[font=\small, align=left, anchor=west] at (3.2,1.6) {shaded: $X$\\$\star$: $\Gamma$};
% right panel: the square X \ Gamma, X, X xor Gamma, X cup Gamma
\begin{scope}[xshift=8.4cm, yshift=1.0cm]
\node[pt] (bot) at (0,-1.3) {$X\setminus\Gamma$};
\node[pt] (lft) at (-1.8,0) {$X$};
\node[pt] (rgt) at (1.8,0) {$X\oplus\Gamma$};
\node[pt] (top) at (0,1.3) {$X\cup\Gamma$};
\node[pt] (up) at (3.9,1.3) {$(X\oplus\Gamma)\cup\Gamma'$};
\draw (bot) -- (lft) -- (top) -- (rgt) -- (bot);
\draw[-{Stealth}, dashed] (rgt.east) to[out=0, in=-90] (up.south);
\node[font=\scriptsize, anchor=north west] at ($(rgt.east)+(0.25,-0.05)$) {if $g(X\oplus\Gamma)>0$};
\end{scope}
\end{tikzpicture}
\caption{Left: a set $X\subseteq A$ (shaded) and a set $\Gamma$ (stars) with one element in each block. Right: $X$ and $X\oplus\Gamma$ have union $X\cup\Gamma$ and intersection $X\setminus\Gamma$. Fixing every coordinate outside $\Gamma$ gives a cube with one free coordinate per block, on which $g=G-F^{(-1)}$ is submodular. If $g(X)<0$, the inequality $g(X)+g(X\oplus\Gamma)\ge g(X\cup\Gamma)+g(X\setminus\Gamma)$ shows that $g$ is nonzero at $X\cup\Gamma$ or $X\setminus\Gamma$, or positive at $X\oplus\Gamma$, from where the change propagates upward. If $g(X)>0$, the change propagates upward from $X$ itself.}
\label{fig:propagation}
\end{figure}
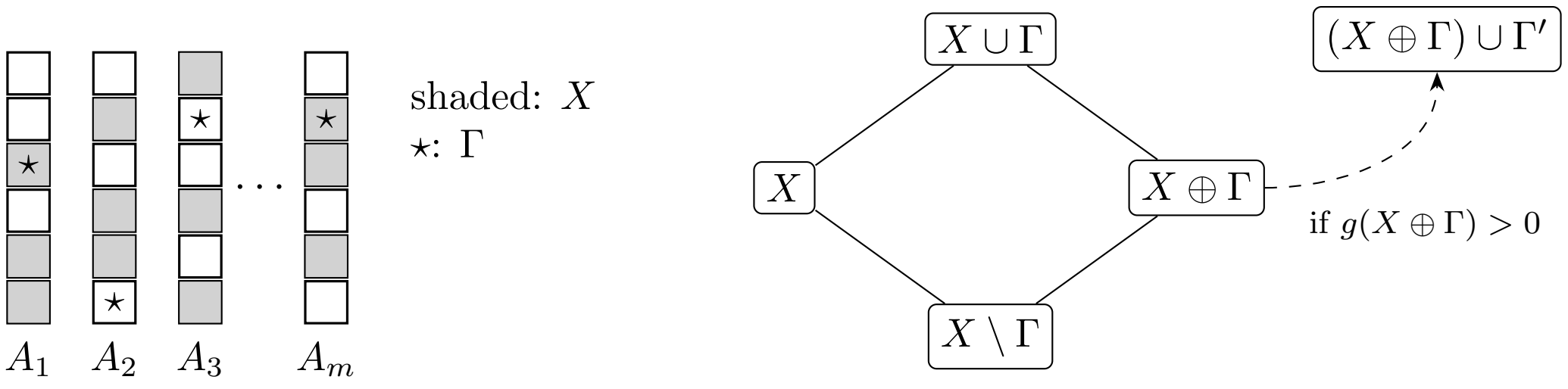

\paragraph{Why the NO functions are far from submodular.}
Let $G$ be submodular.
As in the example, $G$ must differ from $F^{(-1)}$ at one of the four points $(X,Y,\tau)$ of every center $(X,Y)$; say $\tau=01$, so that we work with $Y$ fixed.
Let $\Gamma\subseteq A$ contain exactly one element of each block.
On the subcube in which the coordinates of $\Gamma$ are free and all other coordinates are fixed, $F^{(-1)}(\cdot,Y,01)$ is modular, so $g:=G(\cdot,Y,01)-F^{(-1)}(\cdot,Y,01)$ is submodular there (\cref{lem:closure}).
Restrictions of this kind replace the linear slices of the example (Figure~\ref{fig:propagation}).
 
Suppose $g(X)\neq0$, and that $X$ is \emph{balanced}, meaning that each block $X\cap A_i$ contains between roughly $m/4$ and $3m/4$ elements. If $g(X)>0$, a weighted version of~\cite[Lemma~3.5]{SeshadhriV14} (\cref{clm:weighted-subcube-changes}) shows that $g(X\cup\Gamma)\neq0$ for at least a quarter of the choices of $\Gamma$.
If $g(X)<0$, submodularity applied to $X$ and $X\oplus\Gamma$, whose union is $X\cup\Gamma$ and whose intersection is $X\setminus\Gamma$, gives
\[
    g(X)+g(X\oplus\Gamma)\;\ge\; g(X\cup\Gamma)+g(X\setminus\Gamma).
\]
So $g$ is nonzero at $X\cup\Gamma$ or at $X\setminus\Gamma$, or $g(X\oplus\Gamma)>0$, in which case the positive case applies at $X\oplus\Gamma$.
In every case, $g$ is nonzero at one of $X\cup\Gamma$, $X\setminus\Gamma$ and $(X\oplus\Gamma)\cup\Gamma'$ for at least a quarter of all pairs $(\Gamma,\Gamma')$ (\cref{lem:induced-changes}).
 
Each balanced center therefore forces many changes, but different centers may force the same ones.
To control this overlap, we define the \emph{load} of a set $U$ as the number of triples $(X,\Gamma,\Gamma')$, with $(X, Y)$ a center (recall that $Y$ is fixed), such that $U$ is one of the three candidate sets. % \yynote{Could we say ``with $(X,Y)$ a center'' for fixed $Y$, and specify that $U$ is one of the three candidate sets? This load is independent of $G$, so the resulting distance bound holds simultaneously for all submodular $G$.}
This load is independent of $G$, so the resulting distance bound holds simultaneously for all submodular $G$.
By Cauchy--Schwarz, the number of changed points is at least the square of the number of forced changes, counted with multiplicity, divided by the sum of squared loads (\cref{eq:s-lower-bound}).
Because the centers are random, the loads are close to uniform: the expected sum of squared loads is within a constant factor of its value under perfectly even loads (\cref{lem:expected-load}).
Together with the fact that most centers are balanced (\cref{lem:balanced-center-count}), this shows that $F^{(-1)}$ is $10^{-8}$-far from submodular with probability at least $8/10$.
% \yynote{Could we make the final count explicit across all slices and fixed halves? If $M$ counts forced changes with multiplicity and $N$ is the domain size, a squared-load bound $O(M^2/N)$ gives $\Omega(N)$ distinct changes. \dipi{I can't think of a concise way to add this; it is explained quite well in the actual technical section. If you have a wording in mind, feel free to add it.}}
 
\paragraph{The two roles of the blocks.}
The blocks serve two purposes.
First, as explained above, they let the base components absorb the random terms while remaining sums of functions of single blocks, so that restrictions with one free coordinate per block are modular and changes propagate.
Second, they control the second-moment calculation in \cref{lem:expected-load}.
If each half had $r$ blocks of size $b$, the expected sum of squared loads would exceed its value under even loads by roughly a factor of $(1+3/b)^r$: one factor of $1+O(1/b)$ per block.
This stays bounded when $b=\Omega(r)$.
On the other hand, the lower bound comes from the $r$ random bits in each label and is $2^{\Theta(r)}$.
Taking $r=b=m$ balances the two requirements; it gives $n=2m^2+2$ and a lower bound of $2^{\Theta(m)}=2^{\Theta(\sqrt{n})}$.
 
\paragraph{Why the two distributions are hard to distinguish.}
Slices $00$ and $10$ are the same in both distributions, and the sign enters only through slice $01$, masked by the table.
We couple the two distributions so that they use the same labels, and so that the NO table equals the YES table shifted by $2/2^n$, cyclically, on every address that the algorithm first sees in slice $01$.
The two runs then receive identical answers until the algorithm sees some address in both slice $01$ and slice $11$, that is, until it finds a center, or until a shift wraps around, which has probability $2/2^n$ per address (\cref{lem:indistinguishable}).
Finding a center is hard.
When the algorithm first queries slice $01$ at a set $Y$, the label $\phi_B(Y)$ is uniform and independent of all earlier answers, so each set $X$ that appeared in an earlier slice-$11$ query satisfies $\phi_B(Y)=\pi_A(X)$ with probability $2^{-m}$; queries in slice $11$ are symmetric.
An algorithm that makes $q$ queries therefore finds a center with probability at most $q^2/2^m$ (\cref{lem:center-discovery}), which is small unless $q=2^{\Omega(m)}=2^{\Omega(\sqrt{n})}$.

\section{Preliminaries}
\label{sec:prelims}

\paragraph{Notation.}
For a positive integer $m$, let $[m]:=\{1,\dots,m\}$.
For a finite set $S$, we write $2^S$ for the family of subsets of $S$, and we identify each subset with its incidence vector in $\{0,1\}^S$.
For $x,y\in\{0,1\}^n$, we write $x\le y$ if $x_i\le y_i$ for every $i \in [n]$, and we write $x \vee y$ and $x \wedge y$ for the coordinatewise maximum and minimum; for sets, these are the union and the intersection.
We write $e_i$ for the $i$-th unit vector, and $\mathbf{1}[E]$ for the indicator of an event or statement $E$.
We use bold symbols for random objects and the corresponding nonbold symbols for their realizations.
 
\paragraph{Submodularity and squares.}
We formally define submodularity.
 
\begin{definition}[Submodularity]\label{def:submodular}
    A function $f\colon\{0,1\}^n\to\mathbb{R}$ is \emph{submodular} if
    \[
        f(x)+f(y)\;\ge\; f(x\vee y)+f(x\wedge y)\qquad\text{for all } x,y\in\{0,1\}^n.
    \]
\end{definition}
 
A \emph{square} is given by a point $x\in\{0,1\}^n$ and two coordinates $u\neq v$ with $x_u=x_v=0$.
We write $x^u:=x+e_u$, $x^v:=x+e_v$ and $x^{uv}:=x+e_u+e_v$, and we say that the square with corners $x$, $x^u$, $x^v$ and $x^{uv}$ is \emph{spanned} by $u$ and $v$ at $x$.
The \emph{slack} of the square with respect to $f$ is
\[
    f(x^u)+f(x^v)-f(x)-f(x^{uv}).
\]
The slack is linear in $f$: the slack of $f+h$ is the slack of $f$ plus the slack of $h$.
A function is \emph{modular} if every square has slack $0$.
For example, every function of the form $h(x)=\sum_{i=1}^n h_i(x_i)$ is modular, because it changes by the same amount along opposite sides of every square.
 
\begin{lemma}[Characterizations of submodularity]\label{lem:squares}
    For $f\colon\{0,1\}^n\to\mathbb{R}$, the following are equivalent.
    \begin{enumerate}
        \item[(a)] $f$ is submodular.
        \item[(b)] Every square has nonnegative slack.
        \item[(c)] (Diminishing returns.) For all $x\le y$ and every coordinate $u$ with $x_u=y_u=0$,
        \[
            f(x^u)-f(x)\;\ge\; f(y^u)-f(y).
        \]
    \end{enumerate}
\end{lemma}
 
The equivalence of (a) and (c) is standard (see, e.g., \cite[Section~1]{SeshadhriV14}). We include a short proof of all three for completeness.
 
\begin{proof}
    \emph{(a) implies (b).} Apply the definition to $x^u$ and $x^v$, using $x^u\vee x^v=x^{uv}$ and $x^u\wedge x^v=x$.
     
    \emph{(b) implies (c).} Let $v_1,\dots,v_k$ be the coordinates in which $y$ is $1$ and $x$ is $0$ (none of them is $u$).
    Put $z_0:=x$ and $z_j:=z_{j-1}+e_{v_j}$, so that $z_k=y$.
    For each $j$, the square spanned by $u$ and $v_j$ at $z_{j-1}$ has corners $z_{j-1}$, $z_{j-1}^u$, $z_j$ and $z_j^u$, so its nonnegative slack says
    \[
        f(z_{j-1}^u)-f(z_{j-1})\;\ge\; f(z_j^u)-f(z_j).
    \]
    Chaining these $k$ inequalities gives (c).
     
    \emph{(c) implies (a).} Let $w:=x\wedge y$, and let $a_1,\dots,a_k$ be the coordinates in which $x$ is $1$ and $y$ is $0$.
    Put $w_j:=w+e_{a_1}+\dots+e_{a_j}$ and $y_j:=y+e_{a_1}+\dots+e_{a_j}$, so that $w_0=w$, $w_k=x$, $y_0=y$ and $y_k=x\vee y$.
    For each $j$, we have $w_{j-1}\le y_{j-1}$, and coordinate $a_j$ is $0$ in both, so (c) gives
    \[
        f(w_j)-f(w_{j-1})\;\ge\; f(y_j)-f(y_{j-1}).
    \]
    Summing over $j$ gives $f(x)-f(w)\ge f(x\vee y)-f(y)$, which is (a).
\end{proof}

Next we state two properties of submodular functions that we will use in our proofs. 
\begin{lemma}[Restrictions and modular shifts]\label{lem:closure}
    Let $f\colon\{0,1\}^n\to\mathbb{R}$ be submodular.
    \begin{enumerate}
        \item[(a)] Fixing any set of coordinates gives a submodular function of the remaining coordinates.
        \item[(b)] For every modular function $h$, the function $f+h$ is submodular.
    \end{enumerate}
\end{lemma}
 
\begin{proof}
We use Lemma~\ref{lem:squares}(b). For (a), every square of the smaller cube is also a square of $\{0,1\}^n$. For (b), the slack of $f+h$ equals the slack of $f$ on every square.
\end{proof}
 
\paragraph{Distance and testers.}
A \emph{property} $\mathcal{P}$ is a set of functions $f\colon\{0,1\}^n\to\mathbb{R}$ for each $n$; for example, submodularity is the property consisting of all submodular functions.
 
\begin{definition}[Distance]\label{def:distance}
    For $f,g\colon\{0,1\}^n\to\mathbb{R}$, let
    \[
    \mathrm{dist}(f,g):=\frac{|\{x\in\{0,1\}^n : f(x)\neq g(x)\}|}{2^n}.
    \]
    The \emph{distance} of $f$ to a property $\mathcal{P}$ is $\mathrm{dist}(f,\mathcal{P}):=\min_{g\in\mathcal{P}}\mathrm{dist}(f,g)$.
    For $\varepsilon>0$, we say $f$ is \emph{$\varepsilon$-far} from $\mathcal{P}$ if $\mathrm{dist}(f,\mathcal{P})\ge\varepsilon$.
\end{definition}

Next we define a tester.
 
\begin{definition}[$\varepsilon$-tester]\label{def:tester}
    Let $\mathcal{P}$ be a property and let $\varepsilon\in(0,1)$. An \emph{$\varepsilon$-tester} for $\mathcal{P}$ is a randomized algorithm that is given $n$ and query access to a function $f\colon\{0,1\}^n\to\mathbb{R}$, where each query asks for the value $f(x)$ at a point $x$ of the algorithm's choice. For all sufficiently large $n$, it must satisfy the following.
    \begin{itemize}
    \item If $f\in\mathcal{P}$, it accepts with probability at least $2/3$.
    \item If $f$ is $\varepsilon$-far from $\mathcal{P}$, it rejects with probability at least $2/3$.
    \end{itemize}
    The tester may choose its next query based on the answers to previous queries, that is, it may be \emph{adaptive}, and it may err on both kinds of inputs, that is, it has \emph{two-sided error}.
    Its \emph{query complexity} is the maximum number of queries it makes on functions over $\{0,1\}^n$, as a function of $n$.
\end{definition}
 
To rule out testers with polynomial query complexity, it suffices to prove a superpolynomial lower bound for infinitely many \(n\). % \yynote{Could we explicitly add ``for infinitely many $n$ to the lower-bound theorem and its restatement in the introduction? This matches the intended scope described here; the tester definition can stay as it is. \dipi{Is this really necessary? I think the theorem statement reads well as-is.} \yy{We can just put ``, for infinitely many $n$.'' at the end of the statement, but it's a matter of taste.} \dipi{I added a short paragraph under the theorem statement in section 3 mentioning exactly what we prove. If you prefer the sentence be in the theorem statement, you can make that change and remove this paragraph. I'd still keep the version in the intro as-is.}}
 
\paragraph{Probability tools.}
We use the following standard inequalities.
 
\begin{lemma}[Markov's inequality]\label{lem:markov}
    If $X$ is a nonnegative random variable and $a>0$, then
    \[
        \Pr[X\ge a]\le \frac{\mathbb{E}[X]}{a}.
    \]
\end{lemma}
 
\begin{lemma}[Chebyshev's inequality]\label{lem:chebyshev}
    If $X$ is a random variable and $a>0$, then
    \[
        \Pr \left[ |X-\mathbb{E}[X]| \ge a \right] \; \le \; \frac{\mathrm{Var}[X]}{a^2}.
    \]
\end{lemma}
 
\begin{lemma}[Chernoff bound]\label{lem:chernoff}
    Let $X$ be a sum of independent random variables taking values in $\{0,1\}$, and let $\mu:=\mathbb{E}[X]$. For every $\delta\in(0,1)$,
    \[
        \Pr \left[ |X-\mu| \ge \delta \mu \right] \; \le \; 2e^{-\delta^2\mu/3}.
    \]
\end{lemma}
 
\begin{lemma}[Cauchy--Schwarz inequality]\label{lem:cauchy-schwarz}
    For real numbers $a_1,\dots,a_k$ and $b_1,\dots,b_k$,
    \[
        \left( \sum_{i=1}^k a_ib_i \right)^2 \; \le \; \left( \sum_{i=1}^k a_i^2 \right) \left( \sum_{i=1}^k b_i^2 \right).
    \]
\end{lemma}

\section{Lower Bound}
\label{sec:lb}

We prove the following lower bound. % \yynote{I think it suffices to state the lower bound for infinitely many $n$. Could we add this qualification to the theorem and take $n=2m^2+2$ for sufficiently large integers $m$ in the final proof? \dipi{I'll just add this to the definition of a tester? That keeps the lower bound statement as clean as possible.}}

\begin{theorem}[\cref{thm:main}, restated]
    \label{thm:lb}
    For $\varepsilon \in \left( 0, \frac{1}{10^8} \right)$, every $\varepsilon$-tester for submodularity must make at least $2^{\sqrt{n}/5}$ queries on functions from $\{ 0, 1 \}^n$ to $\mathbb{R}$.
\end{theorem}

As mentioned in \cref{sec:prelims}, it suffices to prove a lower bound for infinitely many input dimensions $n$.
Our construction can also be extended to work for all large enough $n$ by standard padding arguments, but that is outside the scope of this paper. 

We use Yao's minimax principle \cite{Yao77} in the form of \cref{clm:yao} to prove the theorem.
In \cref{sec:construction}, we define our YES and NO distributions.
In \cref{sec:correctness}, we prove that the distributions are correctly supported, that is, every function in the YES distribution is submodular, and functions in the NO distribution are $\frac{1}{10^8}$-far from submodularity with probability at least $\frac{8}{10}$.
%\yynote{Could we state the two guarantees explicitly? Every YES function is submodular, whereas a NO function is far with probability at least $8/10$. ``Correctly supported'' suggests that every NO function is far.}
In \cref{sec:indistinguishability}, we prove that the distributions are hard to distinguish, and complete the proof of \cref{thm:lb}.

Throughout the rest of the section, let $m$ be a large enough integer.

\subsection{Construction}
\label{sec:construction}

Divide the $n$ input coordinates into $2$ {\em action} coordinates and $2m^2$ {\em control} coordinates.
We hide violations along the action coordinates.
Call the subcube induced by fixing the action coordinates a {\em slice}.
Divide the $2m^2$ control coordinates into two {\em halves} $A$ and $B$, each of length $m^2$.
Divide each half into $m$ {\em blocks} $A_1, \dots, A_m$ and $B_1, \dots, B_m$, each of length $m$.
Thus $n = 2m^2 + 2$. % \yynote{Could we briefly explain the two roles of the block structure, here or in the overview? One free coordinate per block makes the relevant restrictions modular, so changes can be propagated. Later, the moment calculation incurs a factor $1+O(1/m)$ per block; over $m$ blocks this remains constant. Together with the $m$ fresh random bits in an address, this explains why the construction uses $\Theta(m^2)$ coordinates. We need not present the detailed constants here or claim that this choice is necessary. \dipi{Yes I'll try to do this in the technical overview; you can leave the footnote here for now.}}
% We identify subsets with their incidence vectors.
We use $\oplus$ for symmetric difference of sets and for bitwise addition modulo two; for bit vectors $u,v$, the quantity $|u\oplus v|$ is their Hamming distance.

For $X \subseteq A$, let $\pi_A(X) \in \{ 0, 1 \}^m$ be defined as $\pi_A(X)_i = |X \cap A_i| \mod 2$.
Similarly define $\pi_B(Y)$ for $Y \subseteq B$.

Choose random maps $\bm{\phi}_A:2^A\to\{0,1\}^m$ and $\bm{\phi}_B:2^B\to\{0,1\}^m$ by sampling all their values independently and uniformly.
Define the {\em addresses} $\bm{\alpha}(X),\bm{\beta}(Y)\in\{0,1\}^{2m}$ by concatenation:
\[
    \bm{\alpha}(X) := (\pi_A(X),\bm{\phi}_A(X)), \qquad
    \bm{\beta}(Y) := (\bm{\phi}_B(Y),\pi_B(Y)).
\]
Define the random set of {\em centers}
\[
    \bm{\mathcal{C}} := \{(X,Y)\in 2^A\times2^B : \bm{\alpha}(X)=\bm{\beta}(Y)\}.
\]
Equivalently, $(X,Y)\in\bm{\mathcal{C}}$ if and only if $\pi_A(X)=\bm{\phi}_B(Y)$ and $\bm{\phi}_A(X)=\pi_B(Y)$.
For the NO distribution, the squares along the action coordinates will violate submodularity precisely at centers.
For every $X\subseteq A$ and $Y\subseteq B$,
\[
    \Pr[(X,Y)\in\bm{\mathcal{C}}]=2^{-2m}=2^{-\Theta(\sqrt n)}.
\]

The YES and NO functions have the following components.
\begin{enumerate}
    \item {\bf Base components.} To ensure submodularity on squares spanned by two control coordinates, define the submodular functions
    \[
        F_A(X, Y) := -4 \sum_{i = 1}^{m} (| X \cap A_i | + |Y|)^2, \qquad
        F_B(X, Y) := -4 \sum_{i = 1}^{m} (| Y \cap B_i | + |X|)^2.
    \]

    \item {\bf Sign component.} Let $\sigma\in\{+1,-1\}$ be the sign distinguishing the two distributions.
    To mask the shift $\sigma/2^n$ in slice $01$, sample a table
    \[
        \bm{\theta}:\{0,1\}^{2m}\to
        \left\{\frac{0}{2^n},\frac{1}{2^n},\dots,\frac{2^n-1}{2^n}\right\}
    \]
    with independent uniform entries, independently of $\bm{\phi}_A$ and $\bm{\phi}_B$.

    \item {\bf Correction components.}
    To ensure submodularity on squares (nonnegative slack) spanned by the action coordinates at non-centers, define
    \[
        \mathbf{H}_A(X, Y) := | \pi_A(X) \oplus \bm{\phi}_B(Y) |, \qquad
        \mathbf{H}_B(X, Y) := | \bm{\phi}_A(X) \oplus \pi_B(Y) |.
    \]
    We also use the linear correction $m^4(|X|+|Y|)$ to ensure submodularity on squares spanned by one action coordinate and one control coordinate. %\yynote{Could we replace the heading ``Balancing components'' by a direct description of $H_A,H_B$ as the two mismatch counts, or simply use their symbols? No balance condition is being imposed here, and ``balanced'' later has a separate precise meaning for sets. The later phrase ``balancing terms'' in the mixed-square proof could likewise name $H_A,H_B$. \dipi{The balance overload is a good point, but at least when I was trying to come up with the construction from scratch, both of these terms got introduced exactly to balance the slacks to ensure submodularity, so I think keeping it this way should be fine.} \yy{But it wasn't clear to me what ``balancing (slack)'' means. Maybe ``correction components''?}}
\end{enumerate}
Define $\mathbf{F}^{(\sigma)}:\{0,1\}^n\to\mathbb{R}$ by
\begin{align*}
    \mathbf{F}^{(\sigma)} (X, Y, 0, 0) &:= F_A(X, Y)\\
    \mathbf{F}^{(\sigma)} (X, Y, 0, 1) &:= F_A(X, Y) - m^4 \cdot (|X| + |Y|) + \bm{\theta}(\bm{\beta}(Y)) + \frac{\sigma}{2^n} + \mathbf{H}_A(X, Y)\\
    \mathbf{F}^{(\sigma)} (X, Y, 1, 0) &:= F_B(X, Y) - m^4 \cdot (|X| + |Y|)\\
    \mathbf{F}^{(\sigma)} (X, Y, 1, 1) &:= F_B(X, Y) - 2 m^4 \cdot (|X| + |Y|) + \bm{\theta}(\bm{\alpha}(X)) - \mathbf{H}_B(X, Y).
\end{align*}
The YES distribution sets $\sigma = +1$, and the NO distribution sets $\sigma = -1$. % \yynote{Could we restore the explanation in the two commented-out sentences below? At a center, the table values and the other terms cancel in the action-square slack, leaving $\sigma/2^n$. This explains how the table can mask the sign in individual answers while preserving the intended violation.}
At a center $(X,Y)\in\bm{\mathcal{C}}$, we have $\bm{\theta}(\bm{\alpha}(X))=\bm{\theta}(\bm{\beta}(Y))$ and $\mathbf{H}_A(X,Y)=\mathbf{H}_B(X,Y)=0$.
In the square spanned by the two action coordinates, the terms involving $F_A,F_B$ and $|X|+|Y|$ cancel as well, leaving slack $\sigma/2^n$.

\subsection{Correctness}
\label{sec:correctness}

To analyze the correctness of our distributions, we require the following characterization of the slack of squares in $F^{(\sigma)}$.

\begin{lemma}
    \label{lem:square-slack}
    When $(X, Y) \in \bm{\cC}$, the slack in squares along the action coordinates is exactly $\frac{\sigma}{2^n}$.
    For all other squares the slack is always nonnegative for large enough $m$.
\end{lemma}

We prove this in \cref{sec:square-slack-proof}. % \yynote{Several references display ``Theorem'' for results of other types. Could we correct the reference names for lemmas, propositions, and claims? For example, \cref{lem:square-slack} refers to Lemma~\ref{lem:square-slack}. \dipi{This might be a ChatGPT hallucination?}}
Completeness follows as an immediate corollary.

\begin{proposition}[Completeness]
    \label{prop:completeness}
    For large enough $m$, the function $\mathbf{F}^{(+1)}$ is submodular.
\end{proposition}
\begin{proof}
    Plugging in $\sigma = +1$ in \cref{lem:square-slack} shows that every square of $\mathbf{F}^{(+1)}$ has nonnegative slack, meaning it is submodular by \cref{lem:squares}.
\end{proof}

The rest of the section is used to prove the following soundness proposition.

\begin{proposition}[Soundness]
    \label{prop:soundness}
    For large enough $m$, the function $\mathbf{F}^{(-1)}$ is $\frac{1}{10^8}$-far from submodularity with probability at least $8/10$.
\end{proposition}
\begin{proof}
    Counting one changed corner per center only gives a bound proportional to the density of centers, whose expectation is $2^{-2m}$.
    We use submodularity on restrictions to force further changes, then bound how often the same input is counted.
    
    Fix realizations $\phi_A,\phi_B,\theta$ and a submodular function $G:\{0,1\}^n\to\mathbb{R}$.
    By \cref{lem:square-slack}, for every center $(X,Y)\in\mathcal{C}$, at least one slice $\tau\in\{00,01,10,11\}$ must satisfy
    $G(X,Y,\tau)\ne F^{(-1)}(X,Y,\tau)$.
    This is because the square has negative slack in $F$ by \cref{lem:square-slack}, while it has nonnegative slack in $G$ by \cref{lem:squares}.
     % \yynote{Could we restore the two sentences commented out at the start of this proof? Counting one changed corner per center only gives a bound on the scale of the center density. Explaining that submodularity forces further changes, whose overlaps we then control, would motivate the cube restrictions and load definition that follow.}
    
    Fix $\tau\in\{00,01\}$ and $Y\subseteq B$, and define $g_{Y,\tau}:2^A\to\mathbb{R}$ by
    \[
        g_{Y,\tau}(X):=G(X,Y,\tau)-F^{(-1)}(X,Y,\tau).
    \]
    In these slices, $F^{(-1)}(X,Y,\tau)$ is a sum of terms, each depending only on $X\cap A_i$ for some $i$.
    We therefore consider sets of free coordinates that select one element from each block:
    \[
        \mathcal{G}_A:=\{\Gamma\subseteq A:|\Gamma\cap A_i|=1\text{ for every }i\in[m]\}.
    \]
    Define $\mathcal{G}_B$ analogously using the $B$ blocks; both families have $m^m$ members.
    For $\Gamma\in\mathcal{G}_A$, fixing all coordinates outside $\Gamma$ gives an $m$-dimensional cube.
    On this cube, each block contributes a function of a single free bit, so the restriction of $F^{(-1)}$ is modular.
    Subtracting this restriction from $G$ preserves submodularity, and hence $g_{Y,\tau}$ is submodular on the cube.
    For $\tau\in\{10,11\}$, we instead fix $X$ and define $g_{X,\tau}:2^B\to\mathbb{R}$ by $g_{X,\tau}(Y):=G(X,Y,\tau)-F^{(-1)}(X,Y,\tau)$.
    The same reasoning applies using $\mathcal{G}_B$.
    
    For $\pi\in\{0,1\}^m$, define the parity class
    \[
        \mathcal{A}_\pi:=\{X\subseteq A:\pi_A(X)=\pi\}.
    \]
    For $Y\subseteq B$, the sets $X$ for which $(X,Y)$ is a center form the family
    \[
        \mathcal{C}|_Y:=\{X\in\mathcal{A}_{\phi_B(Y)}:\phi_A(X)=\pi_B(Y)\}.
    \]
    
    \paragraph{Changes induced by repairing centers.}
    We first define a balanced set.
    % We first show that a nonzero center which is not far from the middle of the hypercube forces many nonzero points across the half $A$.\yynote{A center is a pair $(X,Y)$; it is $g_{Y,\tau}(X)$ that is nonzero. Since the lemma below states the precise claim, could we simply delete this preview sentence?}
    
    \begin{definition}[Balanced set]
        \label{def:balanced-set}
        A set $X \subseteq A$ (resp.\ $Y \subseteq B$) is {\em balanced} if $|X \cap A_i| \in \left[ \frac{m}{4} + 1, \frac{3m}{4} - 1 \right]$ (resp.\ $|Y \cap B_i| \in \left[ \frac{m}{4} + 1, \frac{3m}{4} - 1 \right]$) for all $i \in [m]$.
    \end{definition}
    
    For a balanced $X$, the margin of one ensures that $m/4\le |(X\oplus\Gamma)\cap A_i|\le3m/4$ for every $\Gamma\in\mathcal{G}_A$ and $i\in[m]$.
    We use the following lemma, which is proved in \cref{sec:induced-changes-proof}. % \yynote{Could we include $Y\subseteq B$ and $\tau\in\{00,01\}$ in the lemma's setup? At the start of its deferred proof, please also recall that $g_{Y,\tau}$ is submodular on every cube with one free coordinate per block. These facts are needed again after a substantial gap.}
    
    \begin{lemma}
        \label{lem:induced-changes}
        Let $X \subseteq A$ be balanced, let $Y \subseteq B$ and $\tau \in \{ 00, 01 \}$, and suppose $g_{Y, \tau}(X) \neq 0$.
        For at least $m^{2m}/4$ pairs $(\Gamma,\Gamma')\in\mathcal{G}_A^2$, $g_{Y,\tau}$ is nonzero at one of the following sets.
        \[
            X \cup \Gamma, \qquad X \setminus \Gamma, \qquad (X \oplus \Gamma) \cup \Gamma'.
        \]
    \end{lemma}
    
    \paragraph{Charging centers to changed points.}
    To turn the count in \cref{lem:induced-changes} into a lower bound on the number of changed inputs, we must control how many tuples lead to each input.
    
    \begin{definition}[Load]
        \label{def:load}
        For $Y\subseteq B$ and $U\subseteq A$, define the load $L_Y(U)$ as the number of triples $(X,\Gamma,\Gamma')$ with $X\in\mathcal{C}|_Y$ and $(\Gamma,\Gamma')\in\mathcal{G}_A^2$ for which $U$ is one of the following sets.
        \[
            X \cup \Gamma, \qquad X \setminus \Gamma, \qquad (X \oplus \Gamma) \cup \Gamma'.
        \]
    \end{definition}
    
    Define $L_X(Y)$ analogously by fixing $X$ and using coordinate sets from $\mathcal{G}_B$.
    For each slice $\tau$, assign the corresponding load to each pair $(X,Y)$:
    \[
        L_\tau(X,Y) :=
        \begin{cases}
            L_Y(X), & \tau \in \{00,01\},\\
            L_X(Y), & \tau \in \{10,11\}.
        \end{cases}
    \]
    Let $S\subseteq 2^A\times2^B\times\{0,1\}^2$ be the set of inputs at which $G$ and $F^{(-1)}$ differ, and let
    $\mathcal{C}' := \{(X,Y)\in\mathcal{C} : X,Y \text{ are balanced}\}$.
    The sum of squared loads over the whole domain is
    \[
        L := \sum_{\tau}\sum_{X\subseteq A}\sum_{Y\subseteq B} L_\tau(X,Y)^2
        = 2\sum_{X\subseteq A}\sum_{Y\subseteq B}
          \bigl(L_Y(X)^2+L_X(Y)^2\bigr).
    \]
    Every center in $\mathcal{C}'$ has a changed corner in at least one slice.
    Choose one such slice for each center and apply \cref{lem:induced-changes} in the corresponding half.
    For each center, at least $m^{2m}/4$ pairs of coordinate sets lead to a changed point in the chosen slice, and each pair contributes to the load of that point.
    Consequently,
    \begin{equation}
        \label{eq:s-lower-bound}
        \frac{m^{2m}}{4}|\mathcal{C}'|
        \le \sum_{(X,Y,\tau)\in S} L_\tau(X,Y)
        \le \sqrt{|S|}\sqrt{L},
    \end{equation}
    where the last step is Cauchy--Schwarz (\cref{lem:cauchy-schwarz}).
    Both $|\mathcal{C}'|$ and $L$ depend only on $\phi_A,\phi_B$, so \cref{eq:s-lower-bound} holds simultaneously for all submodular functions $G$.
    Our targets are
    \[
        |\mathcal{C}'| = \Omega(2^{2m^2-2m}),\qquad
        L = O(m^{4m}2^{2m^2-4m}),
    \]
    which together imply $|S|=\Omega(2^{2m^2})$, a constant fraction of the full domain.
    
    \paragraph{Bounding the sum of squared loads.}
    Fix $Y\subseteq B$ and condition on $\bm{\phi}_B=\phi_B$, leaving $\bm{\phi}_A$ random.
    The set $\bm{\mathcal{C}}|_Y$ contains each point of $\mathcal{A}_{\phi_B(Y)}$ independently with probability $\Pr[\bm{\phi}_A(X) = \pi_B(Y)] = \frac{1}{2^m}$.
    Thus the expected size of the set is % \yynote{In the following display, could we write the cardinality of the random family as $\bigl|\bigl(\bm{\mathcal C}|_Y\bigr)\bigr|$? The current bars obscure what is being counted. Please also use bold symbols for the random $\mathbf L_Y$ in the subsequent display, consistently with our notation convention; the same issue occurs with $\mathcal C'$ inside the variance in the balanced-center proof.}
    \[
        \mathbb{E} \left[ \left| \left( \mathcal{C}|_Y \right) \right| \right] = \frac{1}{2^{m}} \cdot \frac{2^{m^2}}{2^m}.
    \]
    So the expected number of triples is $m^{2m} \cdot \frac{2^{m^2}}{2^{2m}}$.
    If these tuples covered the space of $X \subseteq A$ uniformly, then the sum of squared loads would be
    \[
        % L_Y :=
        \sum_{U \subseteq A} \mathbf{L}_Y(U)^2 
        = \sum_{U \subseteq A} \left( \frac{ m^{2m} \cdot \frac{2^{m^2}}{2^{2m}} }{2^{m^2}} \right)^2
        = m^{4m} \cdot 2^{m^2 - 4m}.
        % = \frac{m^{4m}}{2^{2m}} \cdot \mathbb{E} \left[ |\mathcal{C}|_Y| \right].
    \]
    We show that the expected sum of squared loads is within a constant factor of this, and then sum over $Y$ to upper bound $\mathbf{L}$.
    % We bound the expected sum of squared loads uniformly in $\phi_B(Y)$, so that the upper bound can be summed over $Y$ to control $\mathbf{L}$. % \yynote{Could the paragraph heading say ``Bounding the sum of squared loads''? This is the quantity $L$ defined above, rather than the sum of the loads themselves.}

    \begin{lemma}
        \label{lem:expected-load}
        For $m \ge 5$,
        \[
            \mathbb{E}\left[\sum_{U\subseteq A} \mathbf{L}_Y(U)^2\right]
            \le 250 \cdot m^{4m} \cdot 2^{m^2-4m}.
        \]
        Here the expectation is over $\bm{\phi}_A$, with $Y$ and $\phi_B$ fixed.
    \end{lemma}
    We prove \cref{lem:expected-load} in \cref{sec:expected-load-proof}.
    
    Summing \cref{lem:expected-load} over $Y\subseteq B$, taking expectation over $\bm{\phi}_B$, and using the analogous bound with $A$ and $B$ interchanged gives
    \[
        \mathbb{E}[\mathbf{L}]
        \le 4\cdot 2^{m^2}\cdot 250\,m^{4m}2^{m^2-4m}
        =1000\,m^{4m}2^{2m^2-4m}.
    \]
    Applying Markov's inequality (\cref{lem:markov}), with probability at least $\frac{9}{10}$,
    \begin{equation}
        \label{eq:total-load-pr-bound}
        \sqrt{\mathbf{L}}
        \le 100 \cdot m^{2m} \cdot 2^{m^2 - 2m}.
    \end{equation}
    
    \paragraph{Bounding the number of balanced centers.}
    The expected number of centers $|\bm{\mathcal{C}}|$ is $\frac{1}{2^{2m}} \cdot 2^{2m^2}$.
    We show that the number of balanced centers is at least a quarter of this with high probability.
    
    \begin{lemma}
        \label{lem:balanced-center-count}
        For large enough $m$,
        \[
            \Pr \left[ \left| \bm{\mathcal{C}}' \right| \le \frac{1}{4} \cdot \frac{1}{2^{2m}} \cdot 2^{2m^2} \right] \le \frac{1}{10}.
        \]
    \end{lemma}
    We prove \cref{lem:balanced-center-count} in \cref{sec:balanced-center-count-proof}.
    
    \paragraph{Bounding distance to submodularity.}
    Combining these bounds, we are ready to complete the proof of soundness.
    By \cref{eq:s-lower-bound,eq:total-load-pr-bound,lem:balanced-center-count}, with probability at least $8/10$, the following bound holds simultaneously for all submodular functions $G$:
    \begin{equation*}
        \sqrt{|\mathbf{S}|} \ge \frac{m^{2m} \cdot \frac{1}{4} \cdot 2^{2m^2 - 2m}}{4 \cdot 100 \cdot m^{2m} \cdot 2^{m^2 - 2m}} = \frac{2^{m^2}}{1600}
        \quad \implies \quad
        |\mathbf{S}| \ge \frac{1}{256 \times 10^4} \cdot 2^{2m^2} = \frac{1}{1024 \times 10^4} \cdot 2^n.
    \end{equation*}
    Thus the distance to submodularity is at least $\frac{1}{10^8}$ with probability at least $\frac{8}{10}$.
\end{proof}

\subsubsection{Proof of \cref{lem:square-slack}}
\label{sec:square-slack-proof}

We prove \cref{lem:square-slack} for every $m\ge 9$.
Every square is spanned by two action coordinates, by two control coordinates, or by one coordinate of each kind. \cref{clm:action,clm:control,clm:mixed} treat these three cases separately.
 
Recall that a square is given by a point $x$ and two coordinates $u \neq v$ with $x_u = x_v=0$.
Writing $x^u$, $x^v$ and $x^{uv}$ for $x$ with coordinate $u$, coordinate $v$, or both set to $1$, the slack of the square is
\[
    f(x^u) + f(x^v) - f(x) - f(x^{uv}).
\]
We write points as $(X,Y,\tau)$, where $X \subseteq A$ and $Y\subseteq B$, and $\tau$ is the slice.
% , and we write $F$ for $F^{(\sigma)}$.
Adding a control coordinate $u$ means replacing $X$ by $X \cup \{u\}$ if $u \in A$, or $Y$ by $Y \cup \{ u \}$ if $u\in B$.
 
\paragraph{Two action coordinates.}
This is the simplest case.

\begin{claim}\label{clm:action}
    For all $X\subseteq A$ and $Y\subseteq B$, the slack of the square spanned by the two action coordinates at $(X,Y)$ is $\frac{\sigma}{2^n}$ if $(X,Y) \in\mathcal C$, and nonnegative otherwise.
\end{claim}
 
\begin{proof}
The square consists of the four points $(X,Y,\tau)$ with $\tau\in\{00,01,10,11\}$, and its slack is
\[
    F^{(\sigma)} (X,Y,01) + F^{(\sigma)} (X,Y,10) - F^{(\sigma)} (X,Y,00) - F^{(\sigma)} (X,Y,11).
\]
After substituting the definitions, the terms $F_A$ and $F_B$ cancel, and so do the multiples of $m^4(|X|+|Y|)$.
What remains is the expression
\[
    H_A(X,Y) + H_B(X,Y) + \theta(\beta(Y)) - \theta(\alpha(X)) + \frac{\sigma}{2^n}.
\]
 
If $(X,Y)\in\mathcal C$, then $\pi_A(X)=\phi_B(Y)$ and $\phi_A(X)=\pi_B(Y)$, so $H_A(X,Y) = H_B(X,Y)=0$.
Moreover $\alpha(X) = \beta(Y)$, so the two table values cancel, and the slack is exactly $\frac{\sigma}{2^n}$.
 
If $(X,Y) \notin \mathcal C$, then $\pi_A(X) \neq \phi_B(Y)$ or $\phi_A(X) \neq \pi_B(Y)$, so $H_A(X,Y) + H_B(X,Y)\ge 1$.
Every table value lies between $0$ and $1 - \frac{1}{2^n}$, so the difference of the two table values is at least $-1 + \frac{1}{2^n}$.
Finally $\frac{\sigma}{2^n} \ge -\frac{1}{2^n}$.
Adding the three bounds shows that the slack is at least $0$.
\end{proof}
 
\paragraph{Two control coordinates.}
We first compute the slack of the base terms.
 
\begin{claim}\label{clm:base}
    Let $u \neq v$ be control coordinates.
    The slack of $F_A$ at every square spanned by $u$ and $v$ is
    \begin{itemize}
        \item $0$ if $u$ and $v$ lie in different blocks of $A$;
        \item $8$ if $u$ and $v$ lie in the same block of $A$, or if one lies in $A$ and the other in $B$;
        \item $8m$ if both lie in $B$.
    \end{itemize}
    The same holds for $F_B$ with the roles of $A$ and $B$ exchanged.
    In particular, $F_A$ and $F_B$ are submodular.
\end{claim}
 
\begin{proof}
    The $i$-th summand of $F_A$ is $-4 z_i^2$, where $z_i := |X \cap A_i| + |Y|$ is the number of elements of $(X\cap A_i) \cup Y$.
    Suppose that adding $u$ increases $z_i$ by $s \in \{0,1\}$ and adding $v$ increases it by $t \in \{0,1\}$.
    The slack of the summand is then
    \[
        -4 \left( (z_i+s)^2 + (z_i+t)^2 -z_i^2 -(z_i+s+t)^2 \right) = 8st.
    \]
    So the $i$-th summand contributes $8$ if both $u$ and $v$ lie in $A_i\cup B$, and $0$ otherwise.
    Summing over $i$ gives the three values.
    The statement for $F_B$ follows by symmetry.
\end{proof}
 
\begin{claim}\label{clm:control}
    Every square spanned by two control coordinates has nonnegative slack.
\end{claim}
 
\begin{proof}
    Both action coordinates are fixed, so the square lies inside one slice.
    Inside a slice, the linear term $m^4 (|X|+|Y|)$ and the constant $\frac{\sigma}{2^n}$ contribute zero slack, because they change by the same amount along opposite sides of the square.
    Hence:
    \begin{itemize}
        \item in slice $00$, the slack is the slack of $F_A$;
        \item in slice $01$, it is the slack of $F_A$ plus the slack of $R_{01} (X,Y) := \theta(\beta(Y)) + H_A(X,Y)$;
        \item in slice $10$, it is the slack of $F_B$;
        \item in slice $11$, it is the slack of $F_B$ plus the slack of $R_{11}(X,Y) := \theta(\alpha(X)) - H_B(X,Y)$.
    \end{itemize}
    Exchanging the roles of $A$ and $B$ turns $F_A$ into $F_B$ and $R_{01}$ into a table value depending only on $X$ plus $H_B(X,Y)$.
    This differs from $R_{11}$ only in the sign of $H_B$, and the bounds below do not use this sign.
    We therefore treat slices $00$ and $01$; slices $10$ and $11$ are the same with $A$ and $B$ exchanged.
     
    By Claim~\ref{clm:base}, every square in slice $00$ has nonnegative slack. In slice $01$, we add the slack of $R_{01}$ to the slack of $F_A$ from Claim~\ref{clm:base}, in three cases.
     
    \emph{Different halves.}
    Let $u\in A_j$ and $v\in B$, and write $X':=X\cup\{u\}$ and $Y':=Y\cup\{v\}$.
    The slack of $F_A$ is $8$.
    The slack of $R_{01}$ is
    \[
        \bigl(R_{01}(X',Y)-R_{01}(X,Y)\bigr)-\bigl(R_{01}(X',Y')-R_{01}(X,Y')\bigr).
    \]
    The table term $\theta(\beta(\cdot))$ depends only on the second argument, so it cancels inside each bracket.
    Adding $u$ changes the parity of $|X\cap A_j|$ and no other entry of $\pi_A(X)$, so it changes $H_A(X,Y)$ by exactly $+1$ or $-1$, whatever $Y$ is.
    Hence each bracket equals $\pm1$, the slack of $R_{01}$ is at least $-2$, and the total slack is at least $8-2=6$.
     
    \emph{Both in half $A$.}
    Suppose first that $u\in A_j$ and $v\in A_l$ with $j\neq l$.
    The slack of $F_A$ is $0$.
    Since $Y$ is fixed, $\theta(\beta(Y))$ is a constant.
    Moreover, $H_A(X,Y)$ counts the blocks $i$ for which the parity of $|X\cap A_i|$ differs from $\phi_B(Y)_i$, so it is a sum of $m$ terms, the $i$-th depending only on $X\cap A_i$.
    Adding $u$ changes only the $j$-th term and adding $v$ changes only the $l$-th, so the slack of $R_{01}$ is $0$.
    The total slack is $0$.

    Suppose next that $u,v\in A_j$.
    The slack of $F_A$ is $8$.
    As in the previous case, $\theta(\beta(Y))$ is a constant, and only the $j$-th term of $H_A$ changes.
    This term depends only on the parity of $|X\cap A_j|$.
    The corners $X$ and $X \cup \{u,v\}$ have one parity, and the corners $X\cup\{u\}$ and $X\cup\{v\}$ have the other.
    So if the term equals $c\in\{0,1\}$ at $X$, the slack of $R_{01}$ is
    \[
        (1-c) + (1-c) - c - c = 2 - 4c \ge-2,
    \]
    and the total slack is at least $8-2=6$.
     
    \emph{Both in half $B$.}
    Let $u$ and $v$ both lie in $B$, in the same block or not. % \yynote{This includes the same-block case under the heading ``Same half, different blocks.'' Could we organize the cases as one coordinate in each half, both in $A$ (with two block cases), and both in $B$? This would also avoid returning to the $B$ case below.}
    The slack of $F_A$ is $8m$.
    Since $0 \le \theta < 1$ and $0 \le H_A \le m$, every value of $R_{01}$ lies in an interval of length $m+1$, namely $[0,m+1)$.
    The slack of $R_{01}$ is a sum of two of its values minus a sum of two others, so it is greater than $-2(m+1)$.
    The total slack is greater than $8m-2(m+1)=6m-2>0$.
    % If $u$ and $v$ lie in the same block of $B$, the argument above for two coordinates of $B$ applies unchanged.
\end{proof}
 
\paragraph{One action coordinate and one control coordinate.}
We first bound how much the terms of $F^{(\sigma)}$ can grow when one control element is added.
 
\begin{claim}\label{clm:growth}
    Let $(X',Y')$ be obtained from $(X,Y)$ by adding one control element.
    Going from $(X,Y)$ to $(X',Y')$:
    \begin{enumerate}
        \item[(a)] each of $\theta(\alpha(X))$, $\theta(\beta(Y))$ and $\theta(\alpha(X))-\theta(\beta(Y))$ increases by less than $1$;
        \item[(b)] the absolute changes of $H_A$ and $H_B$ add up to at most $m+1$;
        \item[(c)] $F_B-F_A$ increases by less than $8m^3+8m^2$.
    \end{enumerate}
\end{claim}
 
\begin{proof}
    (a) Each table value lies in $[0,1)$, so it increases by less than $1$.
    Adding an element changes $X$ or $Y$ but not both, so at most one of $\theta(\alpha(X))$ and $\theta(\beta(Y))$ changes, and the same bound applies to their difference.
     
    (b) Adding an element of $A$ changes exactly one entry of $\pi_A(X)$, so $H_A$ changes by exactly $+1$ or $-1$. % \yynote{Minor: could we say ``changes by $+1$ or $-1$''? The claim bounds the absolute change; the function does not always increase.}
    It also replaces the label $\phi_A(X)$ by the label of a different set, so $H_B$, which lies between $0$ and $m$, changes by at most $m$.
    Adding an element of $B$ is symmetric.
     
    (c) Adding an element $u\in A_j$ changes only the $j$-th summand of $F_A$, by $-4(2z_j+1)$, where $z_j=|X\cap A_j|+|Y|\le m^2+m-1$.
    It decreases every summand of $F_B$, since $|X|$ grows.
    So $F_B-F_A$ increases by at most $4(2z_j+1)<8m^2+8m$.
    Adding an element of $B$ decreases each of the $m$ summands of $F_A$ by $4(2z_i+1)$, where again $z_i=|X\cap A_i|+|Y|\le m^2+m-1$, so by less than $8m^2+8m$; it also decreases $F_B$. So $F_B-F_A$ increases by less than $m(8m^2+8m)=8m^3+8m^2$.
\end{proof}
 
\begin{claim}\label{clm:mixed}
    For $m\ge 9$, every square spanned by one action coordinate and one control coordinate has positive slack.
\end{claim}
 
\begin{proof}
    Let $(X,Y,\tau)$ be the base point of the square, so its action coordinate is $0$ in $\tau$, and let $\tau'$ be $\tau$ with that coordinate set to $1$. If the control coordinate $u$ lies in $A$, the slack of the square is
    \[
    \left( F^{(\sigma)} (X, Y, \tau') - F^{(\sigma)} (X, Y, \tau) \right)
    - \left( F^{(\sigma)} (X \cup \{ u \}, Y, \tau') - F^{(\sigma)} (X \cup \{ u \}, Y, \tau) \right),
    \]
    and if $u\in B$, the same holds with $Y\cup\{u\}$ in place of $X\cup\{u\}$.
    So it suffices to show that adding one control element strictly decreases the difference $F^{(\sigma)} (X,Y,\tau') -F^{(\sigma)} (X,Y,\tau)$.
    There are four such differences, one for each pair of slices that differ in one action coordinate:
    \begin{align*}
        F^{(\sigma)} (X, Y, 01) - F^{(\sigma)} (X, Y, 00) &= -m^4(|X|+|Y|) + \theta(\beta(Y)) + \frac{\sigma}{2^n} + H_A(X,Y),\\
        F^{(\sigma)} (X, Y, 11) - F^{(\sigma)} (X, Y, 10) &= -m^4(|X|+|Y|) + \theta(\alpha(X)) -H_B(X,Y),\\
        F^{(\sigma)} (X, Y, 10) - F^{(\sigma)} (X, Y, 00) &= -m^4(|X|+|Y|) + F_B(X,Y) - F_A(X,Y),\\
        F^{(\sigma)} (X, Y, 11) - F^{(\sigma)} (X, Y, 01) &= -m^4(|X|+|Y|) + F_B(X,Y) - F_A(X,Y)\\
        &\qquad + \theta(\alpha(X))-\theta(\beta(Y)) -\frac{\sigma}{2^n} - H_A(X,Y) - H_B(X,Y).
    \end{align*}
    Each difference is $-m^4(|X|+|Y|)$ plus a remainder. Adding one control element lowers the first part by exactly $m^4$. Each remainder consists of a constant and at most one part of each kind in Claim~\ref{clm:growth}: table terms, correction terms, and $F_B-F_A$. By Claim~\ref{clm:growth}, the remainder therefore increases by less than
    \[
    1+(m+1)+(8m^3+8m^2)=8m^3+8m^2+m+2.
    \]
    For $m\ge 9$ this is less than $m^4$, because
    \[
    m^4-8m^3-8m^2-m-2 \;\ge\; m^3-8m^2-m-2 \;\ge\; m^2-m-2 \;>\;0,
    \]
    where both steps use $m-8\ge1$. So every difference strictly decreases, and the slack is positive.
\end{proof}

\subsubsection{Proof of \cref{lem:induced-changes}}
\label{sec:induced-changes-proof}

% \begin{proof}[Proof of \cref{lem:induced-changes}]
    We first prove the following fact about submodular functions, which is a weighted version of \cite[Lemma~3.5]{SeshadhriV14}.
    \begin{claim}
        \label{clm:weighted-subcube-changes}
        Let $\ell\ge1$ and let $f : \{ 0, 1 \}^\ell \to \mathbb{R}$ be submodular with $f(0) > 0$, and let $\mathbf{W} \in \{ 0, 1 \}^\ell$ be such that $\Pr[\mathbf{W}_i = 1] = q_i$ independently for all $i \in [\ell]$.
        Then
        \[
            \Pr[f(\mathbf{W}) \neq 0] \ge \min_{i \in [\ell]} \min (q_i, 1 - q_i).
        \]
    \end{claim}
    \begin{proof}
        Induct on $\ell$.
        For $\ell=1$, the assumption $f(0)>0$ gives
        $\Pr[f(\mathbf{W})\ne0]\ge1-q_1\ge\min(q_1,1-q_1)$.
        Now let $\ell\ge2$ and write $q:=\min_{i\in[\ell]}\min(q_i,1-q_i)$.

        \emph{Case $f(e_\ell) \le 0$.}
        Then $f(e_\ell)-f(0)<0$, so by diminishing returns (\cref{lem:squares}(c)), $f(x+e_\ell) - f(x) \le f(e_\ell) - f(0) < 0$ for every $x$ with $x_\ell=0$.
        So on every edge $\{x, x + e_\ell\}$ the two values differ, and at least one of them is nonzero.
        Condition on the first $\ell-1$ coordinates of $\mathbf{W}$.
        Then $\mathbf{W}$ is the lower end of that edge with probability $1 - q_\ell$ and the upper end with probability $q_\ell$, so $\Pr[f(\mathbf{W}) \ne 0] \ge \min(q_\ell,1-q_\ell) \ge q$.
        
        \emph{Case $f(e_\ell)>0$.}
        The restrictions $f_0 := f(\cdot,0)$ and $f_1:=f(\cdot,1)$ are submodular on $\{0,1\}^{\ell-1}$ and positive at the origin.
        By induction, each is nonzero with probability at least $\min_{i \in [\ell-1]}\min(q_i, 1 - q_i) \ge q$ under the first $\ell-1$ coordinates of $\mathbf{W}$.
        Conditioning on $\mathbf{W}_\ell$ expresses $\Pr[f(\mathbf{W})\ne0]$ as a weighted average of these two probabilities, so it is at least $q$.
    \end{proof}

    Recall that the function $g_{Y, \tau}$ is submodular on every cube with one free coordinate per block.
    Suppose $g_{Y,\tau}(X)>0$.
    For this case, it suffices to assume the weaker bounds $|X\cap A_i|\in[m/4,3m/4]$ for all $i$.
    Choose $\bm{\Gamma}$ uniformly from $\mathcal{G}_A$, equivalently by choosing one coordinate from each block independently and uniformly, and put
    \[
        q_i:=\frac{|A_i\setminus X|}{m}\in[1/4,3/4].
    \]
    For each $i$, choose $\mathbf{a}_i$ uniformly from $A_i\setminus X$ and $\mathbf{W}_i\sim\mathrm{Bernoulli}(q_i)$, with all $2m$ choices independent.
    Then $X\cup\bm{\Gamma}$ has the same distribution as $X\cup\{\mathbf{a}_i:\mathbf{W}_i=1\}$.
    For each tuple $a=(a_1,\dots,a_m)$ with $a_i\in A_i\setminus X$, define
    \[
        g_a(W):=g_{Y,\tau}(X\cup\{a_i:W_i=1\}),\qquad W\in\{0,1\}^m.
    \]
    The function $g_a$ is a restriction to a cube with one free coordinate per block, so it is submodular and $g_a(0)>0$.
    By \cref{clm:weighted-subcube-changes}, $\Pr[g_a(\mathbf{W})\ne0]\ge1/4$.
    Averaging over $\mathbf{a}_1,\dots,\mathbf{a}_m$ gives $\Pr[g_{Y,\tau}(X\cup\bm{\Gamma})\ne0]\ge1/4$.
    Since $\Gamma'$ is arbitrary, at least a quarter of all pairs $(\Gamma,\Gamma')$ satisfy the conclusion.

    Next suppose $g_{Y,\tau}(X) < 0$.
    For each $\Gamma\in\mathcal{G}_A$, the sets $X$ and $X\oplus\Gamma$ lie in the cube with free coordinates $\Gamma$ and all other coordinates fixed as in $X$. Submodularity on this cube gives
    \[
        g_{Y,\tau}(X) + g_{Y,\tau}(X \oplus \Gamma) \ge g_{Y,\tau}(X \cup \Gamma) + g_{Y,\tau}(X \setminus \Gamma).
    \]
    If $g_{Y,\tau}(X \cup \Gamma)$ or $g_{Y,\tau}(X \setminus \Gamma)$ is nonzero, then all choices of $\Gamma'$ count.
    Otherwise, $g_{Y,\tau}(X \oplus \Gamma) > 0$.
    Since $X$ is balanced, $|(X\oplus\Gamma)\cap A_i|\in[m/4,3m/4]$ for every $i$. These are the weaker bounds used in the positive case, so at least a quarter of the choices $\Gamma'\in\mathcal{G}_A$ satisfy $g_{Y,\tau}((X\oplus\Gamma)\cup\Gamma')\ne0$.
    \hfill\qed
% \end{proof}

\subsubsection{Proof of \cref{lem:expected-load}}
\label{sec:expected-load-proof}

% \begin{proof}[Proof of \cref{lem:expected-load}]
    Recall that $Y$ and $\phi_B$ are fixed, and the expectation is taken over $\bm{\phi}_A$. 
    Separate the load according to the operation that produces $U$. % \yynote{Could we begin by recalling that $Y$ and $\phi_B$ are fixed and the expectation is over $\bm{\phi}_A$? This conditioning was stated in the lemma, but is needed again here after the intervening proofs.}
    For $U\subseteq A$, let $\mathbf{L}_1(U)$, $\mathbf{L}_2(U)$, and $\mathbf{L}_3(U)$ count triples $(X,\Gamma,\Gamma')$, with $X\in\bm{\mathcal{C}}|_Y$ and $(\Gamma,\Gamma')\in\mathcal{G}_A^2$, satisfying
    \[
        X\cup\Gamma=U,\qquad X\setminus\Gamma=U,\qquad (X\oplus\Gamma)\cup\Gamma'=U,
    \]
    respectively.
    A triple counted by $\mathbf{L}_Y(U)$ may satisfy more than one of these conditions, so
    \[
        \mathbf{L}_Y(U)^2
        \le \bigl(\mathbf{L}_1(U)+\mathbf{L}_2(U)+\mathbf{L}_3(U)\bigr)^2
        \le 3\sum_{j=1}^3\mathbf{L}_j(U)^2.
    \]
    We bound the expected sum of squared loads $\mathbb E[\sum_U\mathbf L_j(U)^2]$ for $j \in [3]$ by counting pairs of coordinates within each block. %\yynote{Could we name the quantity as $\mathbb E[\sum_U\mathbf L_j(U)^2]$ and say ``pairs of coordinates in each block''? $\mathbf L_j$ is the load function, not the sum of its squares, and the block variables $\Gamma_i,\Gamma_i'$ are elements of $A_i$, not coordinate sets.}
    % Complementation will then give the bound for $\mathbf{L}_2$.

    \paragraph{Expanding the squared loads.}
    Write $\pi:=\phi_B(Y)$ for the prescribed parity vector.
    For $X\in\mathcal{A}_\pi$, let $\mathbf{I}_X=\mathbf{1}[X\in\bm{\mathcal{C}}|_Y]$.
    These indicators are independent Bernoulli variables with mean $\frac{1}{2^m}$.
    For $X,U\subseteq A$, define
    \begin{align*}
        \gamma_1(X,U)
        &:=\bigl|\{(\Gamma,\Gamma')\in\mathcal{G}_A^2:X\cup\Gamma=U\}\bigr|,\\
        \gamma_2(X,U)
        &:=\bigl|\{(\Gamma,\Gamma')\in\mathcal{G}_A^2:X\setminus\Gamma=U\}\bigr|,\\
        \gamma_3(X,U)
        &:=\bigl|\{(\Gamma,\Gamma')\in\mathcal{G}_A^2:(X\oplus\Gamma)\cup\Gamma'=U\}\bigr|.
    \end{align*}
    Thus, for all $j\in[3]$,
    \[
        \mathbf{L}_j(U) = \sum_{X \in \mathcal{A}_\pi} \mathbf{I}_X \cdot \gamma_j(X,U).
    \]
    Since $\mathbf{I}_X$ takes values in $\{0,1\}$, we have $\mathbb{E}[\mathbf{I}_X^2] = \frac{1}{2^m}$.
    For $X\ne X'$, independence gives $\mathbb{E}[\mathbf{I}_X\mathbf{I}_{X'}] = \frac{1}{2^{2m}}$.
    Expanding the square and separating the terms with $X = X'$ from those with $X\ne X'$ yields
    \begin{align}
        \mathbb{E} [ \mathbf{L}_j(U)^2 ]
        &= \frac{1}{2^m} \cdot \sum_{X\in\mathcal{A}_\pi}\gamma_j(X,U)^2
        + \frac{1}{2^{2m}} \cdot \sum_{\substack{X,X'\in\mathcal{A}_\pi\\X\ne X'}}\gamma_j(X,U)\gamma_j(X',U)
        \notag\\
        \label{eq:load-second-moment}
        &\le \frac{1}{2^m} \cdot \sum_{X \in \mathcal{A}_\pi} \gamma_j(X,U)^2
        + \frac{1}{2^{2m}} \cdot \left( \sum_{X \in \mathcal{A}_\pi} \gamma_j(X,U) \right)^2.
    \end{align}
    The inequality includes the diagonal terms in the second sum, all of which are nonnegative.

    \paragraph{Splitting the count by blocks.}
    To compute the two sums over $X$ in \cref{eq:load-second-moment}, write $X_i=X\cap A_i$, $U_i=U\cap A_i$, and
    \[
        \mathcal{A}_\pi^{(i)}:=\{X_i\subseteq A_i:|X_i|\equiv\pi_i\pmod 2\}.
    \]
    For $X_i,U_i\subseteq A_i$, define the corresponding counts within block $i$ by
    \begin{align*}
        \gamma_1^{(i)}(X_i,U_i)
        & := \left| \{ (\Gamma_i, \Gamma_i') \in A_i^2 : X_i \cup \{ \Gamma_i \} = U_i \} \right|,\\
        \gamma_2^{(i)}(X_i,U_i)
        & := \left| \{ (\Gamma_i, \Gamma_i') \in A_i^2 : X_i \setminus \{ \Gamma_i \} = U_i \} \right|,\\
        \gamma_3^{(i)}(X_i,U_i)
        & := \left| \{ (\Gamma_i, \Gamma_i') \in A_i^2 : (X_i \oplus \{ \Gamma_i \}) \cup \{ \Gamma_i' \} = U_i \} \right|.
    \end{align*}
    Each choice of one pair $(\Gamma_i, \Gamma_i')$ per block specifies a pair $(\Gamma,\Gamma') \in \mathcal{G}_A^2$, so
    \[
        \gamma_j(X,U)=\prod_{i=1}^m\gamma_j^{(i)}(X_i,U_i),
    \]
    for all $j \in [3]$.
    For each $U_i$, we need both the sum of these counts and the sum of their squares:
    \[
        s_j^{(i)}(U_i) := \sum_{X_i\in\mathcal{A}_\pi^{(i)}} \gamma_j^{(i)} (X_i,U_i),
        \qquad
        t_j^{(i)}(U_i) := \sum_{X_i\in\mathcal{A}_\pi^{(i)}} \gamma_j^{(i)} (X_i,U_i)^2.
    \]
    Since the choices of $X_i\in\mathcal{A}_\pi^{(i)}$ can be combined freely across blocks,
    \[
        \sum_{X \in \mathcal{A}_\pi} \gamma_j (X,U)   = \prod_{i=1}^m s_j^{(i)}(U_i),
        \qquad
        \sum_{X \in \mathcal{A}_\pi} \gamma_j (X,U)^2 = \prod_{i=1}^m t_j^{(i)}(U_i).
    \]
    Summing \cref{eq:load-second-moment} over $U$ and then summing separately over each $U_i\subseteq A_i$ gives
    \begin{align}
        \mathbb{E} \left[ \sum_{U \subseteq A} \mathbf{L}_j(U)^2 \right]
        &\le \sum_{U\subseteq A} \left(
        \frac{1}{2^m} \cdot \prod_{i=1}^m t_j^{(i)}(U_i)
        + \frac{1}{2^{2m}} \cdot \prod_{i=1}^m s_j^{(i)}(U_i)^2\right)\notag\\
        \label{eq:load-per-operation}
        &= \frac{1}{2^m} \cdot \prod_{i=1}^m\left(\sum_{U_i\subseteq A_i}t_j^{(i)}(U_i)\right)
        + \frac{1}{2^{2m}} \cdot \prod_{i=1}^m\left(\sum_{U_i\subseteq A_i}s_j^{(i)}(U_i)^2\right).
    \end{align}
    We now compute $s_j^{(i)}(U_i)$ and $t_j^{(i)}(U_i)$. %, and then sum over $U_i$.

    \begin{claim}
        \label{clm:s-and-t-count-per-block}
        Fix $i \in [m]$ and let $w := |U_i|$.
        The following is true.
        
        \begin{tabular}{lll}
             \vspace{1em}
             1. &
             $\displaystyle s_{1}^{(i)} (U_i) = mw,$ &
             $\displaystyle t_{1}^{(i)} (U_i) = \begin{cases}
                m^2 w^2, &w \equiv \pi_i \pmod 2,\\
                m^2 w, &w \not\equiv \pi_i \pmod 2.
            \end{cases}$\\
            \vspace{1em}
             2. &
            $\displaystyle s_{2}^{(i)} (U_i) = m(m-w),$ &
            $\displaystyle t_{2}^{(i)} (U_i) = \begin{cases}
                m^2 (m - w)^2, &w \equiv \pi_i \pmod 2,\\
                m^2 (m - w), &w \not\equiv \pi_i \pmod 2.
            \end{cases}$\\
            3. &
            $\displaystyle s_{3}^{(i)} (U_i) = mw,$ &
            $\displaystyle t_{3}^{(i)} (U_i) = \begin{cases}
                w (m + 2w - 2), &w \equiv \pi_i \pmod 2,\\
                m w^2, &w \not\equiv \pi_i \pmod 2.
            \end{cases}$\\
        \end{tabular}
    \end{claim}

    \begin{proof}
        The first two cases follow from nearly identical arguments.
        \begin{enumerate}
            \item The condition $X_i \cup \{ \Gamma_i \} = U_i$ requires $\Gamma_i \in U_i$ and either $X_i = U_i$ or $X_i = U_i \setminus \{ \Gamma_i \}$.
            The coordinate $\Gamma_i'$ is arbitrary, giving $m$ choices in either case.
            If $w \equiv\pi_i\pmod 2$, only $X_i = U_i$ has the required parity, and its count is $mw$.
            Otherwise, there are $w$ possible sets $X_i$, each with count $m$.
            Hence
            \[
                s_1^{(i)}(U_i)=mw,\qquad
                t_1^{(i)}(U_i)=
                \begin{cases}
                    (mw)^2, & w\equiv\pi_i\pmod 2,\\
                    w\,m^2, & w\not\equiv\pi_i\pmod 2.
                \end{cases}
            \]

            \item An analogous argument proves this by observing that $X_i \setminus \{ \Gamma_i \} = U_i$ if and only if $\Gamma_i \in A_i \setminus U_i$ and either $X_i = U_i$ or $X_i = U_i \cup \{ \Gamma_i \}$, thus replacing the $w$ term with $(m-w)$.
            % The same arguments hold except with $\Gamma_i \in A_i \setminus U_i$.

            \item We have
            \[
                (X_i \oplus \{ \Gamma_i \}) \cup \{ \Gamma_i' \}=U_i
                \quad\Longleftrightarrow\quad
                \Gamma_i' \in U_i \ \text{and} \ X_i \oplus \{ \Gamma_i \} \in \{ U_i, U_i \setminus \{ \Gamma_i' \} \}.
            \]
            Suppose first that $w \not \equiv\pi_i\pmod 2$.
            Then $X_i \oplus \{ \Gamma_i \}$ has the same parity as $U_i$, so it must equal $U_i$.
            Thus $X_i = U_i \oplus \{ \Gamma_i \}$ gives $m$ distinct choices of $X_i$, each with $w$ choices of $\Gamma_i'$.
            Consequently,
            \[
                s_3^{(i)}(U_i)=mw,\qquad t_3^{(i)}(U_i)=mw^2.
            \]
            Now suppose that $w\equiv\pi_i\pmod 2$.
            In this case $X_i \oplus \{ \Gamma_i \} = U_i \setminus \{ \Gamma_i' \}$, or equivalently
            \[
                X_i = ( U_i \setminus \{ \Gamma_i' \}) \oplus \{ \Gamma_i \}, \qquad \Gamma_i' \in U_i.
            \]
            The cases $\Gamma_i = \Gamma_i'$, $\Gamma_i \in U_i \setminus \{ \Gamma_i' \}$, and $\Gamma_i \notin U_i$ give the following counts.
            \begin{center}
                \renewcommand{\arraystretch}{1.3}
                \begin{tabular}{lcc}
                    Form of $X_i$ & Number of such sets & $\gamma_3^{(i)}(X_i,U_i)$\\
                    \hline
                    $X_i=U_i$ & $1$ & $w$\\
                    $X_i\subseteq U_i$, $|U_i\setminus X_i|=2$ & $\binom{w}{2}$ & $2$\\
                    $|U_i\setminus X_i|=|X_i\setminus U_i|=1$ & $w(m-w)$ & $1$
                \end{tabular}
            \end{center}
            In the second row, the two missing elements can be assigned to $\Gamma_i$ and $\Gamma_i'$ in either order.
            Summing the counts in the table, and then their squares, yields the desired result.
            \begin{align*}
                s_3^{(i)}(U_i)
                &=w+2\binom{w}{2}+w(m-w)=mw,\\
                t_3^{(i)}(U_i)
                &=w^2+4\binom{w}{2}+w(m-w)=w(m+2w-2).
                \qedhere
            \end{align*}
            % Combining the two parity cases, we have
            % \[
            %     s_3^{(i)}(U_i)=mw,\qquad
            %     t_3^{(i)}(U_i)=
            %     \begin{cases}
            %         w(m+2w-2), & w\equiv\pi_i\pmod 2,\\
            %         mw^2, & w\not\equiv\pi_i\pmod 2.
            %     \end{cases}
            %     \qedhere
            % \]
        \end{enumerate}
    \end{proof}

    \paragraph{Summing over the subsets of a block.}
    We next bound the two block sums in \cref{eq:load-per-operation}.

    \begin{claim}
        \label{clm:s-and-t-sum-over-sets}
        For $m \ge 4$ and all $j \in [3]$, the following is true.
        \[
            \sum_{U_i \subseteq A_i} s_j^{(i)} (U_i)^2 = \left( 1 + \frac{1}{m} \right) m^4 \cdot 2^{m-2}, \qquad
            \sum_{U_i \subseteq A_i} t_j^{(i)} (U_i) \le \left( 1 + \frac{3}{m} \right) m^4 \cdot 2^{m-3}.
        \]
    \end{claim}
    \begin{proof}
        For $m\ge3$ and either parity $p \in\{0,1\}$, exactly $2^{m-2}$ subsets $U_i$ of parity $p$ contain a specified coordinate, and exactly $2^{m-3}$ contain two specified distinct coordinates.
        Indeed, after fixing those one or two coordinates, half the assignments of the remaining coordinates have the required parity.
        Expanding
        \[
            w^2
            = \left( \sum_{\Gamma_i \in A_i} \mathbf{1} [\Gamma_i \in U_i] \right)^2
            = \sum_{\Gamma_i \in A_i} \mathbf{1} [\Gamma_i \in U_i]
            + \sum_{\substack{\Gamma_i, \Gamma_i' \in A_i \\ \Gamma_i \neq \Gamma_i'}} \mathbf{1} [\Gamma_i, \Gamma_i' \in U_i],
        \]
        we obtain
        \begin{align}
            \sum_{\substack{U_i\subseteq A_i\\w\equiv p\pmod 2}}w
            &=m \cdot 2^{m-2},\notag\\
            \label{eq:block-parity-moments}
            \sum_{\substack{U_i\subseteq A_i\\w\equiv p\pmod 2}}w^2
            &=m \cdot 2^{m-2} + m(m-1) \cdot 2^{m-3}
            =m(m+1) \cdot 2^{m-3}.
        \end{align}
        Adding the second identity for the two parities gives
        \begin{equation}
            \label{eq:block-square-sum}
            \sum_{U_i\subseteq A_i}w^2=m(m+1) \cdot 2^{m-2}.
        \end{equation}
        Use the values of $s_j^{(i)}$ and $t_j^{(i)}$ from \cref{clm:s-and-t-count-per-block}.
        Since $s_1^{(i)}(U_i)=s_3^{(i)}(U_i)=mw$, for $j\in\{1,3\}$ we get
        \[
            \sum_{U_i\subseteq A_i}s_j^{(i)}(U_i)^2
            =m^2 \cdot \sum_{U_i\subseteq A_i}w^2
            =\left(1+\frac1m\right) \cdot m^4 \cdot 2^{m-2}.
        \]
        For $t_1^{(i)}$, separating the two parities and using \cref{eq:block-parity-moments} gives
        \begin{align*}
            \sum_{U_i\subseteq A_i}t_1^{(i)}(U_i)
            &=m^2 \cdot \left(
            \sum_{\substack{U_i\subseteq A_i\\w\equiv\pi_i\pmod 2}}w^2
            +\sum_{\substack{U_i\subseteq A_i\\w\not\equiv\pi_i\pmod 2}}w\right)\\
            &=m^2\left(m(m+1) \cdot 2^{m-3} + m \cdot 2^{m-2}\right)\\
            &=\left(1+\frac3m\right) \cdot m^4 \cdot 2^{m-3}.
        \end{align*}
        For $s_2^{(i)}$ and $t_2^{(i)}$, the exact same bounds hold by summing over the set $A_i \setminus U_i$ instead.
        This is because $U_i \mapsto A_i \setminus U_i$ is a bijection, and \cref{eq:block-parity-moments} holds for both parities.
        For $t_3^{(i)}$, the two parity cases both satisfy $t_3^{(i)}(U_i)\le2mw^2$: when $w\ge1$, we have $m+2w-2\le2mw$, and when $w=0$ both sides vanish.
        Using \cref{eq:block-square-sum}, we get
        \begin{align*}
            \sum_{U_i\subseteq A_i}t_3^{(i)}(U_i)
            &\le2m \cdot \sum_{U_i\subseteq A_i}w^2\\
            &=\left(1+\frac1m\right) \cdot m^3 \cdot 2^{m-1}\\
            &\le\left(1+\frac3m\right) \cdot m^4 \cdot 2^{m-3},
        \end{align*}
        where the last inequality follows from $4(m+1)\le m(m+3)$ for $m\ge4$.
    \end{proof}
    
    \paragraph{Substituting the block sums.}
    % We have shown, for $j\in\{1,3\}$ and every $i\in[m]$, that
    % \[
    %     \sum_{U_i\subseteq A_i}t_j^{(i)}(U_i)\le\left(1+\frac3m\right) m^4 2^{m-3},
    %     \qquad
    %     \sum_{U_i\subseteq A_i}s_j^{(i)}(U_i)^2=\left(1+\frac1m\right)m^4 2^{m-2}.
    % \]
    % Substituting these bounds into \cref{eq:load-per-operation} yields
    Substitute the bounds from \cref{clm:s-and-t-sum-over-sets} into \cref{eq:load-per-operation}.
    \begin{align}
        \mathbb{E}\left[\sum_{U\subseteq A}\mathbf{L}_j(U)^2\right]
        &\le 2^{-m}\left[\left(1+\frac3m\right)m^4 2^{m-3}\right]^m
        +2^{-2m}\left[\left(1+\frac1m\right)m^4 2^{m-2}\right]^m\notag\\
        &=m^{4m}2^{m^2-4m}
        \left[\left(1+\frac3m\right)^m+\left(1+\frac1m\right)^m\right]\notag\\
        &\le(e^3+e)m^{4m}2^{m^2-4m}\notag\\
        \label{eq:operation-load-bound}
        &\le25\,m^{4m}2^{m^2-4m}.
    \end{align}
    Here we used $(1+3/m)^m\le e^3$ and $(1+1/m)^m\le e$.
    % The bound holds for every prescribed parity vector $\pi$.
    % Finally, for $\mathbf{L}_2$, take complements in $A$:
    % \[
    %     X\setminus\Gamma=U
    %     \quad\Longleftrightarrow\quad
    %     (A\setminus X)\cup\Gamma=A\setminus U.
    % \]
    % Complementation maps each parity class bijectively to another and preserves independent selection with probability $2^{-m}$.
    % Reindexing the sum over $U$ by $A\setminus U$ and applying the bound for $\mathbf{L}_1$ in \cref{eq:operation-load-bound} therefore gives
    % \[
    %     \mathbb{E}\left[\sum_{U\subseteq A}\mathbf{L}_2(U)^2\right]
    %     \le25\,m^{4m}2^{m^2-4m}.
    % \]
    Combining this bound with $\mathbf{L}_Y(U)^2\le3\sum_{j=1}^3\mathbf{L}_j(U)^2$ proves the lemma, since $9 \cdot 25 \le 250$.
    \hfill\qed
% \end{proof}

\subsubsection{Proof of \cref{lem:balanced-center-count}}
\label{sec:balanced-center-count-proof}

% \begin{proof}[Proof of \cref{lem:balanced-center-count}]
    Call a pair $(X, Y) \in 2^A \times 2^B$ {\em balanced} if both $X$ and $Y$ are balanced.
    We first show that for large enough $m$, the total number of balanced pairs is at least half of the hypercube $\{ 0, 1 \}^{2m^2}$. % \yynote{Could we say that a balanced point $(X,Y)$ means that both $X$ and $Y$ are balanced? Before applying Chernoff, please introduce $\mathbf{X}$ and $\mathbf{Y}$ as independent uniform subsets of $A$ and $B$, and use the bold symbols consistently in the probability displays.}

    \begin{claim}
        \label{clm:balanced-point-count}
        For large enough $m$, the number of balanced pairs is at least $\frac{2^{2m^2}}{2}$.
    \end{claim}
    \begin{proof}
        Let $\bm{X}$ and $\bm{Y}$ be independent uniform subsets of $A$ and $B$ respectively.
        For $m \ge 20$, if $\bm{X}$ is not balanced then $\bigl| |\bm{X} \cap A_i| - \frac{m}{2} \bigr | > \frac{m}{5}$ for some $i$; the analogous statement holds for $\bm{Y}$.
        By the Chernoff Bound (\cref{lem:chernoff}), for every $i \in [m]$,
        \[
            \Pr \left[ \left| |\bm{X} \cap A_i| - \frac{m}{2} \right| \ge \frac{m}{5} \right] \le 2e^{-m/200},
        \]
        and analogously for $\bm{Y}$.
        Taking a union bound over all $2m$ blocks,
        \[
            \Pr[\bm{X} \text{ or } \bm{Y} \text{ is not balanced}]
            \le 4m e^{-m/200}\le\frac12,
        \]
        for sufficiently large $m$.
        Thus the number of balanced pairs is at least $\frac{2^{2m^2}}{2}$.   
    \end{proof}
    
    % Chernoff's inequality and a union bound over the $2m$ blocks therefore give, for independent uniform subsets $\mathbf{X}\subseteq A$ and $\mathbf{Y}\subseteq B$,
    % \[
    %     \Pr[\mathbf{X} \text{ or } \mathbf{Y} \text{ is not balanced}]
    %     \le 4m e^{-m/200}\le\frac12
    % \]
    % for sufficiently large $m$.

    Now we bound the number of balanced centers.
    Since each pair is a center with probability $\frac{1}{2^{2m}}$, the mean $\mu:=\mathbb{E}[|\bm{\mathcal{C}}'|]$ satisfies
    \[
        \mu \ge \frac{1}{2^{2m}} \cdot \frac{2^{2m^2}}{2}.
    \]
    The indicators of membership in $\bm{\mathcal{C}}'$ for $(X,Y)$ and $(X',Y')$ are independent unless $X=X'$ or $Y=Y'$.
    There are at most $2 \cdot2^{3m^2}$ ordered pairs sharing one of these subsets, and each covariance is at most $1$.
    Bounding each indicator's variance by its mean (because they are Bernoulli random variables) yields
    \[
        \operatorname{Var} \left[ |\bm{\mathcal{C}}'| \right] \le \mu+2^{3m^2+1}.
    \]
    Chebyshev's inequality (\cref{lem:chebyshev}) thus gives
    \begin{align*}    
        \Pr \left[ |\bm{\mathcal{C}}'| \le \frac{1}{2} \cdot \frac{1}{2^{2m}} \cdot \frac{2^{2m^2}}{2} \right]
        \le \Pr \left[ |\bm{\mathcal{C}}'| \le \frac{1}{2} \cdot \mu \right]
        &\le \frac{4  \Var \left[ |\mathcal{C}'| \right]}{\mu^2} \notag \\
        &\le \frac{4}{\mu} + \frac{2^{3m^2 + 3}}{\mu^2} \notag \\
        &\le \frac{1}{2^{2m^2 - 2m - 3}} + \frac{1}{2^{m^2 - 4m - 5}} \le \frac{1}{10},
    \end{align*}
    % Since $2^{2m^2-2m-2}\le\mu/2$, Chebyshev's inequality gives
    % \begin{equation*}
    %     % \label{eq:balanced-centers-pr-bound}
    %     \Pr\left[ |\bm{\mathcal{C}}'| \le \frac{1}{2} \cdot \frac{1}{2^{2m}} \cdot \frac{2^{2m^2}}{2} \right]
    %     \le \frac{4\operatorname{Var}(|\bm{\mathcal{C}}'|)}{\mu^2}
    %     \le 2^{-2m^2+2m+3}+2^{-m^2+4m+5}
    %     \le\frac{1}{10}
    % \end{equation*}
    for sufficiently large $m$.
    \hfill\qed
% \end{proof}

\subsection{Indistinguishability}
\label{sec:indistinguishability}

Let $\cA$ be a deterministic algorithm making $q$ queries.
We bound the difference between the acceptance probabilities of $\cA$ for the distributions $\mathbf{F}^{(+1)}$ and $\mathbf{F}^{(-1)}$. % \yynote{Lemma~\ref{lem:indistinguishable} bounds the absolute difference between the two acceptance probabilities, rather than a specified success probability. Could we use that description here and before the lemma? Writing $\Pr[\mathcal{A}(\mathbf{F}^{(\sigma)})=1]$ would also distinguish the random function from its distribution. In the next sentence, the additive shift is $\sigma/2^n$, not $\sigma$.}

The sign term $\frac{\sigma}{2^n}$ enters $\mathbf{F}^{(\sigma)}$ only through slice $01$, where it is added to $\bm{\theta}(\bm{\beta}(Y))$.
Slice $11$ also uses the table $\bm{\theta}$ on $\bm{\alpha}(X)$, and the other two slices are the same in both distributions.
We say that algorithm $\cA$ {\em discovers a center} if it makes queries $(X', Y, 01)$ and $(X, Y', 11)$ such that $(X, Y) \in \bm{\mathcal{C}}$, that is, $\bm{\alpha}(X) = \bm{\beta}(Y)$.

We first bound the probability that $\cA$ discovers a center. % \yynote{Could we specify that the probability bound holds for either sign $\sigma$, over the random choices defining $\mathbf F^{(\sigma)}$? This identifies the experiment used in the next lemma.}
\begin{lemma}
    \label{lem:center-discovery}
    For both distributions $\mathbf{F}^{(+1)}$ and $\mathbf{F}^{(-1)}$,
    $
        \displaystyle \Pr [\cA \text{ discovers a center}] \le \frac{q^2}{2^m}.
    $
\end{lemma}
\begin{proof}
   A query in slice $01$ uses the address $\bm{\beta}(Y)$, a query in slice $11$ uses $\bm{\alpha}(X)$, and queries in slices $00$ and $10$ use neither.
   The answers depend on $\bm{\phi}_B$ only through $\bm{\phi}_B(Y)$ for sets $Y$ in slice-$01$ queries, and on $\bm{\phi}_A$ only through $\bm{\phi}_A(X)$ for sets $X$ in slice-$11$ queries.
   Suppose $\mathcal A$ first discovers a center at its $t$-th query, say $(X',Y,01)$.
   Then $Y$ does not appear in an earlier slice-$01$ query, since otherwise the center would have been discovered earlier, so $\bm{\phi}_B(Y)$ is uniform given the first $t-1$ answers.
   A center $(X,Y)$ with $X$ from an earlier slice-$11$ query requires $\bm{\phi}_B(Y)=\pi_A(X)$, which has probability $\frac{1}{2^m}$ for each of the at most $t-1$ such sets.
   A slice-$11$ query is symmetric, with $\bm{\phi}_A(X)$ fresh.
   Summing over $t\le q$ gives at most $\frac{q^2}{2^m}$.
\end{proof}
% \begin{proof}
%     Every query $(X, Y, \tau)$ that $\cA$ makes uses either $\alpha(X)$ or $\beta(Y)$ (or neither).
%     Assume WLOG that the next query made by $\cA$ is in slice $01$, that is, it uses $\beta(Y)$.
%     Note that $\beta(Y) = \alpha(X')$ for an arbitrary $X'$ only if $\phi_B(Y) = \pi_A(X')$, which happens with probability $\frac{1}{2^m}$.
%     Thus, by a union bound, the probability that $\cA$ discovers a center with this query is at most $\frac{q}{2^m}$.
%     Taking a union bound over all queries yields the desired result.
% \end{proof}

Next we bound the difference between the acceptance probabilities of $\cA$ for the two distributions.
\begin{lemma}
    \label{lem:indistinguishable}
    $
        \displaystyle \left| \Pr \left[ \mathcal{A} \left( \mathbf{F}^{(+1)} \right) = 1  \right] - \Pr \left[ \mathcal{A} \left( \mathbf{F}^{(-1)} \right) = 1 \right] \right| \le \frac{q^2}{2^m} + \frac{2q}{2^n}.
    $
\end{lemma}
\begin{proof}
    Run $\mathcal A$ on $\mathbf{F}^{(+1)}$ and $\mathbf{F}^{(-1)}$ with the same $\bm{\phi}_A, \bm{\phi}_B$, drawing each table entry the first time it is needed.
    For a new address, draw $\bm{u}$ uniformly from $\{0,\frac1{2^n},\dots,\frac{2^n-1}{2^n}\}$.
    The YES run uses $\bm{u}$.
    The NO run uses $\bm{u}$ if the address is first needed in slice $11$, and $\bm{u} + \frac{2}{2^n}$, taken cyclically, if it is first needed in slice $01$.
    After the first center discovery or wraparound, continue each run by sampling previously unassigned table entries independently and uniformly.
    % ; once the runs differ, it continues with fresh entries. 
    A cyclic shift is a bijection, so both tables are uniform.
    Until $\mathcal A$ discovers a center, no address is needed in both slices, and the two runs receive the same answers: in slice $01$ both answer $\bm{u}+\frac1{2^n}$ plus the same remaining terms, unless the shift wraps around, which has probability $\frac2{2^n}$ for each of the at most $q$ addresses.
    So the outputs differ with probability at most $\Pr[\mathcal A\text{ discovers a center}]+\frac{2q}{2^n}$, and \cref{lem:center-discovery} completes the proof.
    % \yynote{Could we specify the coupling after the first center discovery or wraparound? Once answers differ, the adaptive query sequences may diverge. We can continue each run with independent uniform entries at previously unassigned addresses, retaining all entries already assigned. Stating this continuation would make clear that each run has the intended marginal distribution. Before either failure event, the choice of cyclic permutation is determined by the past, so applying it to a fresh uniform entry preserves the required independence within each table. \dipi{I can add an explanation for this but (a) I feel like this is implicit, and (b) it needlessly complicates a fairly ``simple'' section of the lower bound.} \yy{Agreed; one sentence should suffice: ``After the first center discovery or wraparound, continue each run by sampling previously unassigned table entries independently and uniformly.''}}
\end{proof}
% \begin{proof}
%     Assume $\cA$ does not discover a center.
%     Then the addresses used for queries in the $01$ slice and $11$ slice are all distinct.
%     For $\mathbf{F}^{(+1)}$, the value $\bm{\theta} (\bm{\beta}(Y)) + \frac{\sigma}{2^n}$ is distributed uniformly over $\left\{\frac{1}{2^n},\frac{2}{2^n},\dots,\frac{2^n}{2^n}\right\}$, and for $\mathbf{F}^{(-1)}$ over $\left\{\frac{-1}{2^n},\frac{0}{2^n},\dots,\frac{2^n-2}{2^n}\right\}$.
%     These distributions are identical unless $\frac{-1}{2^n}$ or $\frac{2^n}{2^n}$ is sampled, which happens with probability $\frac{2}{2^n}$.
%     A union bound over all $q$ queries and \cref{lem:center-discovery} yields
%     \[
%         \left| \Pr_{f \sim \mathbf{F}^{(+1)}} \left[ \mathcal{A}(f) = 1  \right] - \Pr_{f \sim \mathbf{F}^{(-1)}} \left[ \mathcal{A}(f) = 1 \right] \right|
%         \le \Pr [\cA \text{ discovers a center}] + \frac{2q}{2^n}.        
%         \le \frac{q^2}{2^m} + \frac{2q}{2^n}. \qedhere
%     \]
% \end{proof}

We are now ready to complete the proof of \cref{thm:lb}.

\begin{proof}[Proof of \cref{thm:lb}]
    We use Yao's minimax principle as stated in \cref{clm:yao}.
    Fix $\varepsilon \in \left( 0, \frac{1}{10^8} \right)$ and $n = 2m^2 + 2$ for sufficiently large $m \in \mathbb{N}$. % \yynote{Could we instead take $n=2m^2+2$ for sufficiently large integers $m$? This proves the intended lower bound for infinitely many $n$, so no padding is needed.}
    By \cref{prop:completeness,prop:soundness}, the distributions $\mathbf{F}^{(+1)}$ and $\mathbf{F}^{(-1)}$ satisfy the conditions of \cref{clm:yao} with $\eta = \frac{2}{10}$.
    For a deterministic algorithm $\cA$ making
    \[
        q \le 2^{\frac{\sqrt{n}}{5}} \le \frac{2^{\frac{\sqrt{n}}{4}}}{10} \le \frac{2^{\frac{\sqrt{\frac{n-2}{2}}}{2}}}{10} = \frac{2^{\frac{m}{2}}}{10}
    \]
    queries, \cref{lem:indistinguishable} says that the acceptance probability of $\cA$ for the two distributions differs by at most $\frac{1}{10}$ (for large enough $n$), which is less than $\frac{1}{3} - \frac{2}{10}$.
    Thus \cref{clm:yao} yields the desired lower bound.
\end{proof}

\section*{AI Disclosure}

The authors used GPT-6 Astra to assist with the lower bound construction.
The authors then tried to independently come up with their own construction in an attempt to simplify it.
The proofs are entirely human generated.
The authors also used GPT-6 Astra and Claude Opus 5.5 for proofreading and clarity improvements to the final draft. 

\bibliographystyle{alpha}
\bibliography{references}

\appendix

\section{Yao's Minimax Principle}

We use the following form of Yao's minimax principle \cite{Yao77}.
 
\begin{claim}[Yao's principle]\label{clm:yao}
    Let $\mathcal{P}$ be a property of functions over the Boolean hypercube, let $\varepsilon, \eta \in(0,1)$, and let $n \in \mathbb{N}$.
    Let $\mathcal{D}^{(+1)}$ and $\mathcal{D}^{(-1)}$ be distributions over functions on the domain $\{ 0, 1 \}^n$,
    such that $F^{(+1)} \sim \mathcal{D}^{(+1)}$ is always in $\mathcal{P}$ and $F^{(-1)} \sim \mathcal{D}^{(-1)}$ is $\varepsilon$-far from $\mathcal{P}$ with probability at least $1 - \eta$.
    Suppose that every deterministic algorithm $\mathcal{A}$ that makes at most $q$ queries satisfies
    \[
        \left|
            \Pr_{F^{(+1)} \sim \mathcal{D}^{(+1)}} [\mathcal{A} \left( F^{(+1)} \right) = 1]
            - \Pr_{F^{(-1)} \sim \mathcal{D}^{(-1)}} [\mathcal{A} \left( F^{(-1)} \right) = 1]
        \right| < \frac 1 3 - \eta.
    \]
    Then every $\varepsilon$-tester for $\mathcal{P}$ makes more than $q$ queries on functions over $\{ 0, 1 \}^n$.
\end{claim}
\begin{proof}
    Suppose that $\cA$ is an $\varepsilon$-tester for $\mathcal{P}$ that makes at most $q$ queries on functions over $\{ 0, 1 \}^n$, and let $R$ denote its random coins.
    For every fixed value $r$ of $R$, the algorithm $\cA_r$ that runs $\cA$ with coins $r$ is deterministic and makes at most $q$ queries.
    Every $F^{(+1)}$ in the support of $\mathcal{D}^{(+1)}$ is in $\mathcal{P}$, so $\Pr_{F^{(+1)} \sim \mathcal{D}^{(+1)},R} \left[ \cA_R \left( F^{(+1)} \right) = 1 \right]\ge 2/3$.
    Every $F^{(-1)}$ that is $\varepsilon$-far from $\mathcal{P}$ is accepted with probability at most $1/3$, so $\Pr_{F^{(-1)} \sim \mathcal{D}^{(-1)}, R} \left[ \cA_R \left( F^{(-1)} \right) = 1 \right] \le \eta + 1/3$.
    Hence
    \[
        \mathbb{E}_R \left[ 
            \Pr_{F^{(+1)} \sim \mathcal{D}^{(+1)}} [\cA_R \left( F^{(+1)} \right) = 1]
            - \Pr_{F^{(-1)} \sim \mathcal{D}^{(-1)}} [\cA_R \left( F^{(-1)} \right) = 1]
        \right] \ge \frac 1 3 - \eta,
    \]
    so some fixed value $r$ satisfies the same inequality, and $\cA_r$ contradicts the assumption.
\end{proof}

\end{document}

%% file: header.tex
\usepackage[letterpaper, portrait, margin=1in]{geometry}
\usepackage[hypertexnames=false,colorlinks=true,linkcolor=Mahogany,citecolor=ForestGreen,pagebackref=true]{hyperref}
\usepackage{algorithm}
\usepackage[noend]{algpseudocode}
\usepackage{url}
\usepackage{thmtools,thm-restate}
\usepackage{amsmath,amssymb,amsthm}
\usepackage[noabbrev,capitalise,nameinlink]{cleveref}
\usepackage{mathtools}
\usepackage{xspace}
\usepackage{verbatim}
\usepackage{mathrsfs}
\usepackage[usenames,dvipsnames,svgnames,table]{xcolor}
\usepackage{pgf}
\usepackage[dvipsnames]{xcolor}
\usepackage{todo}
\usepackage{tabularx}
\usepackage{enumitem}
\usepackage{derivative}
\usepackage{bm}
\usepackage{multirow}
\usepackage{diagbox}
\usepackage{nicematrix}
\usepackage{parskip}
\usepackage{adjustbox}
\usepackage{tikz}
\usepackage{tikz-3dplot}
\usepackage{transparent}
\usepackage{subcaption}
\usepackage[most]{tcolorbox}

\usepackage{dsfont}

\newcommand*\ie{i.\kern.1em e., }
\newcommand*\eg{e.\kern.1em g., }
\newcommand*\cf{c.\kern.1em f.\ }
\newcommand*\almev{a.\kern.1em e.\ }

\theoremstyle{plain}
\newtheorem{theorem}{Theorem}[section]
\newtheorem{lemma}[theorem]{Lemma}

\newtheorem{proposition}[theorem]{Proposition}
\newtheorem{claim}[theorem]{Claim}

\theoremstyle{definition}
\newtheorem{definition}[theorem]{Definition}

\crefname{claim}{Claim}{Claims}
\crefname{fact}{Fact}{Facts}
\crefname{observation}{Observation}{Observations}

\tcbset{colframe=white,boxrule=0pt,frame hidden,parbox=false}
\tcbset{
    theoremstyle/.style={ colback=blue!7, },
    definitionstyle/.style={ colback=orange!10, },
    notestyle/.style={ colback=gray!8, }
}

\tcolorboxenvironment{restatable}{theoremstyle}
\tcolorboxenvironment{theorem}{theoremstyle}
\tcolorboxenvironment{lemma}{theoremstyle}
\tcolorboxenvironment{fact}{theoremstyle}
\tcolorboxenvironment{proposition}{theoremstyle}
\tcolorboxenvironment{claim}{theoremstyle}
\tcolorboxenvironment{corollary}{theoremstyle}

\tcolorboxenvironment{definition}{definitionstyle}
\tcolorboxenvironment{assumption}{definitionstyle}
\tcolorboxenvironment{conjecture}{notestyle}
\tcolorboxenvironment{observation}{notestyle}
\tcolorboxenvironment{remark}{notestyle}
\newcommand{\ignore}[1]{}

\DeclareMathOperator{\Var}{Var}
\newcommand{\cA}{\ensuremath{\mathcal{A}}}

\newcommand{\cC}{\ensuremath{\mathcal{C}}}

\usepackage{complexity}
\newlang{\NB}{NB}
\newlang{\DB}{DB}
\newlang{\DBW}{DBW}
\newlang{\DBWR}{DBWR}
\newlang{\DBWP}{DBWP}

\usetikzlibrary{positioning, arrows, arrows.meta}
\usetikzlibrary{arrows.meta, decorations.pathreplacing, calc, shapes.geometric}
\usetikzlibrary{patterns}
\usetikzlibrary{backgrounds}